\documentclass[preprint,preprintnumbers,amsmath,amssymb]{revtex4}

\usepackage{graphicx}% Include figure files
\usepackage{dcolumn}% Align table columns on decimal point
\usepackage{bm}% bold math
\begin{document}

\title{ The two dimensional  local density of states  of  a   
Topological Insulator with   an edge  dislocation}

\author{ D. Schmeltzer}

\affiliation{Physics Department, City College of the City University of New York \\
New York, New York 10031}

%\vspace{0.2 in}

\begin{abstract}

We  investigate  the effect of a crystal  edge dislocation on the metallic surface of a  Topological Insulator.  The edge dislocation gives rise to torsion which  the electrons experience as a  spin  connection.  As a result    the electrons propagate along  confined two dimensional regions  and circular   contours. Due to the edge dislocations the parity symmetry is violated resulting in a current measured  by the in-plane component of the spin on the surface.
  The tunneling  density of states  for Burger vectors in the $y$ direction  is maximal along  the $x$ direction . The evidence of the enhanced tunneling density of states  can be verified   with the help of the  scanning tunneling  technique.

\end{abstract}

%\pacs{73.23.Ra }% PACS, the Physics and Astronomy
                             % Classification Scheme.

                              %display desired
\maketitle
%----------------------------------------------------------------------
\textbf{I Introduction}

Seldom a new state of matter is predicted.  Even less often  is it  unambiguously observed in the laboratory in the spectacular way  was  done with HgTe quantum wells \cite{Konig}. Indeed, this novel two-dimensional (2D) topological insulator (TI) shows a quantized charge conductance, similar to that observed in Quantum Hall effect, without the need to apply a magnetic field. This result triggered further theoretical and experimental work that resulted in the discovery of metallic surface states in the insulating alloy $Bi_{x}Sb_{1-x}$, the first three-dimensional topological insulator. In this case, the conducting surface states are formed by topological effects that render the electrons traveling on such surfaces insensitive to scattering by impurities.
The theoretical foundations of this phenomena are  based on  the topology of the  Brillouin zone 
 \cite{Volkov,Haldane,Jansen,Kaplan,Kreutz,Mele,Kane,Fu,Spinorbit, More,ZhangField,Ludwig,Nakahara,dnova}.
The theoretical predictions have  been confirmed  experimentally for the   three dimensional ($3D$) Topological Insulators ($TI$) $Bi_{2}Se_{3}$, $Bi_{2}Te_{3}$  and $Bi_{1-x}Sb_{x}$.  The electronic band structure of these crystals  is time reversal invariant    obeying  Kramer's theorem $T^{2}=-1$  and have  a single Dirac cone which lies in a gap \cite{Volkov,Hasan,Discovery,Konig}. At the boundary  of the $3D$ $TI$,  one obtains a $2D$  surface with an odd number of chiral edge  excitations  coined     $helical$ $liquid$   \cite{Wu}  and  realized  experimentally in the  two-dimensional    $CdTe/HgTe/CdTe$  quantum wells \cite{Konig,BernewigZhang}.
The dissipationless surface states are believed to be quantum-protected by the bulk insulator.
A variety of  transport experiments suggest that the conductivity of a  $3D$   $TI$   contains a significant  metallic contribution \cite{Ando}. 
  Scanning tunneling microscopy  ($STM$)  and  transmission electron spectroscopy ($TEM$ )  of the chiral metallic boundary  show that  crystal  defects modify  the density of states.
The momentum-resolved Landau spectroscopy of the Dirac surface state in $Bi_{2}Se_{3}$  \cite{Hanaguri} reveals,   in addition to the Landau spectroscopy,  triangular-shaped structure   caused by the presence of vacancies of $Bi$ at the $Se$ sites.  
These experiments suggest that   the  two dimensional surface of the $TI$  is sensitive to defects  and geometry  and therefore, the   quantization rules are expected  to be modified, and probably  are revealed through the new  Berry indices  $\gamma\neq\frac{1}{2}$ \cite{McDonalds}.
(In a recent Shubnikov-de Haas oscillation experiment performed on the $TI$  $Bi_{2}Se_{3}$ \cite{McDonalds} the authors  show  magnetic oscillations which correspond to the   Landau level quantization $E_{n}=v_{F}\sqrt{2(n+\gamma)\hbar e B}$ with  the   Berry phase  index which is different from $\gamma=\frac{1}{2}$.)
The appearance of  the new indices  $\gamma\neq\frac{1}{2}$ might be due to topology of the defects or/and interactions.  
   Recently \cite{Balatsky}  the effect of the  $\delta^{(2)}(\vec{r})$  impurity potential  on the  metallic surface of the $TI$  has been shown to modify   the local density of states. 
  The authors in ref.  \cite{Ran} have proposed that  crystal  dislocations  generate   protected zero modes which give rise to perfect   metallic conduction. This idea has been used to explain the thermoelectric transport in the $3D$ $TI$ \cite{Sinova}.

From the theory of quantum  crystal \cite{kosevich} three  primary types of  topological defects   are known : $edge$ $dislocations$,   $screw$ $dislocations$ and  $mixed$    $dislocations$. The strength  of the  dislocations  is measured in units of  of the $Burger$ vectors   which  corresponds to the shortest lattice translation in the crystal.  Experimentally,  the dislocations are seen as dark lines in the lighter central regions of the  transmission electron micrograph.

The  dislocations  induce crystal  stresses which  are characterized by  the Burger vector,  shear modulus and  Poisson's ratio \cite{kosevich,Nelson}. As a result, the lattice vibrations are complicated and the electron-phonon interaction plays a crucial role in inducing  new dynamical  effects.
We will ignore the dynamical effects and concentrate on the static effects caused  by the elastic strain field on the electrons. In a crystal we have the neutrality condition,    namely   the sum  of the Burger vectors must be zero.  We  consider first a single  dislocation (one assumes that a  second dislocation with an opposite Burger vector  is located far away from the first dislocation)  and in the  second stage we generalize the results to an even number of  dislocations.

Thirty two  years ago the propagation of electrons in the presence  of  a dislocated  crystal was  investigated  \cite{Kawamura}. Those calculations, performed with the    Schr\"odinger equation,  revealed the possibility for  interesting    electronic  transport.   For a  screw dislocation, the Schr\"odinger equation is equivalent to the Aharonov-Bohm problem in two dimensions \cite{Aharonov} which hints  that the physics of persistent currents might play a crucial role .
In graphene,   different topological defects such as  crystal  $disclinations$ have been modeled as a vortex field  \cite{alberto,Vozmediano}.

We start our discussion  with the     Peierls model \cite{Peierls} which   is  an approximation for   the  edge dislocations without  torsion . This model  gives   results  which are in  agreement with those  obtained  by  \cite{Ran} for  the $3D$ $TI$.  For this  model we find that at the boundary of the dislocation  a zero mode exist  and    confines the electrons to the dislocation line.

Unfortunately, the Peierls model represents a over- simplification of the  reality.  The model is lacking torsion  which is an important property for dislocations. A complete    description of   the edge dislocation  must include  the $torsion$ tensor .  This tensor  generates    spin  connections \cite{Pnueli,Green,Birrell,dnova} which are     controlled by the $Burger$ vector.
Using the complete  description  of the edge dislocation  we find  that  the electronic excitations are confined  to a two dimensional region   and to a set of   circular contours.  Such structure  can be studied  using  the formulation introduced in ref.  \cite{Jaffe,Exner, Costa}.   The  circular contours have a   radius $R_{g}(n)$ $n=\pm1,\pm2,\pm3,...$  which are  determined by the strength of the Burger vector.
Comparing the   results  obtained from the model with torsion to   the one obtained from the    \textbf{Peierls} model, we find  that  the \textbf{torsion} destroys   the   zero mode  state,  but due to  the Parity violation   the system is only weakly affected by  backscattering.  

The contents  of this paper is as follows: In chapter II  we present the chiral model for the boundary surface of the  $TI$. Chapter III is devoted to   the edge dislocation.  We devote a short discussion to the  Peierls model  and present an extended  derivation for  the effect of dislocations on the $TI$. Chapter IV is devoted to the solution of the wave function in the presence of an edge dislocation. Chapter V presents the explicit wave functions for the two dimensional region  and circular contours.  Chapter VI presents the the computation of the tunneling density of states induced by the edge  dislocation. 
In chapter  VII we show that the violation of the parity symmetry by the   edge dislocation generates  a  current   which has an in-plane  spin component. We suggest that this current  which confirms the presence of the edge dislocation might be    measured using \textbf{Magnetic Force Microscopy} techniques.  Chapter VIII is devoted to conclusions.

\vspace{0.2 in}
 
\textbf{II-The chiral metal  - the boundary surface  of the three dimensional  Topological Insulator}

\vspace{0.2 in}

The low energy  Hamiltonian for the bulk $3D$ $TI$ in the $Bi_{2}Se_{3}$ family was  shown to behave on the boundary  surface  (the $x,y$- plane) as  a two dimensional   chiral metal  \cite{nature} which is similar  the Rashba  \cite{Rashba}  model.
\begin{eqnarray}
&&H= \int\,d^{2}r \Psi^{\dagger}(\vec{r})[h^{T.I}(x,y)-\mu]\Psi(\vec{r})]\equiv \hbar v_{F}\int\,d^{2}r \Psi^{\dagger}(\vec{r})[i\sigma^{1}\partial_{y} -i\sigma^{2}\partial_{x}-\mu ]\Psi(\vec{r})\nonumber\\&&
\end{eqnarray}
$h^{T.I}(x,y)=\hbar v_{F}[i\sigma^{1}\partial_{y} -i\sigma^{2}\partial_{x}]$
is the   chiral Dirac Hamiltonian in the first quantized language. $v_{F}\approx  5\cdot10^{5} \frac{m}{sec}$ is the Fermi velocity, $\sigma$ is the Pauli matrix describing the electron spin and $\mu$ is the chemical potential  measured relative to the Dirac $\Gamma$ point.
The  Hamiltonian  for the  two dimensional surface $L\times L$ describes well the excitations smaller than the bulk gap of the $3D$ $TI$ at $0.3$ $eV$. Moving away from  the $\Gamma$ point, the Fermi velocity becomes momentum dependent; therefore,  we will introduce a  momentum  cut off  $\Lambda $  to restrict the validity of the Dirac model.
The chiral Dirac model in the Bloch representation   takes the form: $h=\hbar v_{F}(  \vec{K}\times\vec{\sigma})\cdot\hat{z}\equiv  \hbar v_{F}(-\sigma^{1}k_{y}+\sigma^{2}k_{x})$  where periodic boundary conditions imply   $k_{x}=\frac{2 \pi }{L}m_{x}$ , $m_{x}=0,\pm1,\pm2,..$ and  $k_{y}=\frac{2 \pi }{L}m_{y}$ , $m_{y}=0,\pm1,\pm2,..$. The  eigen-spinors for this Hamiltonian are :
$|u(\vec{K})>=[|u_{\uparrow}(\vec{K})>,|u_{\downarrow}(\vec{K})>]^{T}=|\vec{K}>\otimes[1,i e^{i\chi(k_{x},k_{y})}]^{T}$  where  $\chi(k_{x},k_{y})=tan^{-1}(\frac{k_{y}}{k_{x}})$  is the spinor phase and  $\epsilon=\hbar v_{F}\sqrt{k^{2}_{x}+k^{2}_{y}}$ is the eigenvalue  for particles .  For holes we have the eigenvalue   $\epsilon=-\hbar v_{F}\sqrt{k^{2}_{x}+k^{2}_{y}}$  and eigenvectors  $|v(\vec{K})>=[|v_{\uparrow}(\vec{K})>,|v_{\downarrow}(\vec{K})>]^{T}= |\vec{K}>\otimes[-1, i e^{i\chi(k_{x},k_{y})}]^{T}$.
The \textbf{chirality}  operator is defined  in terms of the chiral phase  $\chi(k_{x},k_{y})$: 
\begin{equation}
(\vec{\sigma}\times \frac{\vec{K}}{ |\vec{K}|})\cdot\hat{z}\equiv \sin[\chi(k_{x},k_{y})]\sigma^{1}-\cos[\chi(k_{x},k_{y})]\sigma^{2}
\label{eqchirality}
\end{equation}
The chirality operator  takes the eigenvalue  $-$ (counter-clockwise) for  particles 
$[\sin(\chi(k_{x},k_{y}))\sigma^{1}-\cos(\chi(k_{x},k_{y}))\sigma^{2}]|\vec{K}>\otimes[1,i e^{i\chi(k_{x},k_{y})}]^{T}=-|\vec{K}>\otimes[1,i e^{i\chi(k_{x},k_{y})}]^{T}$
and  $+$ (clockwise) for holes
$[\sin(\chi(k_{x},k_{y}))\sigma^{1}-\cos(\chi(k_{x},k_{y}))\sigma^{2}]|\vec{K}>\otimes[-1,i e^{i\chi(k_{x},k_{y})}]^{T}=|\vec{K}>\otimes[-1,i e^{i\chi(k_{x},k_{y})}]^{T}$.
The Hamiltonian $h^{T.I}(\vec{K})=\hbar v_{F}(-\sigma^{1}k_{y}+\sigma^{2}k_{x})$ is time reversal invariant   $h^{T.I.}(-\vec{K})=T h^{T.I.}(\vec{K})T^{-1}$   ($T=(-i\sigma_{y})K_{0}$   is the time reversal operator and $K_{0}$ is the conjugation operator). At   $\vec{K}=0$  we have  $h^{T.I.}(\vec{K}=0)=T h^{T.I.}(\vec{K}=0)T^{-1}$  therefore  the eigenstate is a \textbf{Kramer  degenerate}. The eigenstate  $|u(\vec{K})>$ obeys $T^{2}|u(\vec{K})>=-|u(\vec{K})>$. The states $|u(\vec{K})>$ and $T|u(\vec{K})>$ are orthogonal to each other. Since $T|u(\vec{K})>\propto |u(-\vec{K})>$ the property  $T^{2}=-1$  guarantees  that backscattering is prohibited.

\vspace{0.2 in}
 
\textbf{III-The   edge dislocation on the  metallic Surface of the three dimensional  Topological Insulator}

\vspace{0.2 in}

\textbf{A-The   Peierls model for the edge dislocation}

\vspace{0.2 in}

We start our presentation with the  Peierls  phenomenogical model  \cite{Peierls} which might be useful for a large number of dislocations.
An edge dislocation in two dimensions can be formulated in the following way: We cut the the two dimensional crystal into two halves ($y<0$ and $y>0$). After cutting the crystal along the line $y=0$, two  surfaces $[x,y<0]$ and  $[x,y>0]$ have been created. We glue back the two halves  by translating  by   minus one  half of the lattice spacing  along $x$ for $y<0$ and one half  along $x$ for $y>0$. This procedure creates a dislocation  along the line  $y=0$  with a Burger vector along $y$ , $B^{(2)}$. We introduce a distribution of  dislocations   $\rho(x,y)$ then the quantity $\rho(x,y)dx dy$ represents the Burger vector passing  through the area $dx dy$.
In order to understand the  effect of   the dislocations on the chiral  fermions, we rewrite the two dimensional model  on a lattice. On a lattice the model is characterized   by  the hopping  matrix elements      $t=t_{0}$ (for opposite spins)  and $t'=t'_{0}$ ( for parallel spins). 
In the absence  of the Peierls   deformation the  lattice model takes the  form:
\begin{eqnarray}
&&H^{L}= 
\sum_{[x=na,y=ma]} \Psi^{\dagger}(\vec{r})[\sigma^{x}(-i t)\hat{\partial}_{y} +\sigma^{y}(i t)\hat{\partial}_{x}]\Psi(\vec{r})+\Psi^{\dagger}(\vec{r})[\sigma^{z}(-2t'+M)] \Psi(\vec{r})\nonumber\\&&+ (-2t')\Psi^{\dagger}(\vec{r})\hat{\nabla}_{x}\Psi(\vec{r})+(-2t')\Psi^{\dagger}(\vec{r})\hat{\nabla}_{y}\Psi(\vec{r})\nonumber\\&&
\end{eqnarray}
where $\hat{\partial}_{y}$, $\hat{\partial}_{x}$  represents  the $discrete$ lattice derivatives and the action of the discrete operators $\hat{\nabla}_{x}$ $\hat{\nabla}_{y}$  on the field operator is: $\hat{\nabla}_{x}\Psi(\vec{r})\equiv \Psi(\vec{r}+a_{x})+\Psi(\vec{r}-a_{x})-2\Psi(\vec{r}) $ and $\hat{\nabla}_{y}\Psi(\vec{r})\equiv \Psi(\vec{r}+a_{y})+\Psi(\vec{r}-a_{y})-2\Psi(\vec{r})$.  When  $2t'_{0}=M$ we obtain the   chiral model given in eq.$(1)$. 
\begin{equation}
H^{L}\approx\sum_{[x=na,y=ma]} \Psi^{\dagger}(\vec{r})[\sigma^{x}(-it)\hat{\partial}_{y} +\sigma^{y}(it)\hat{\partial}_{x}]\Psi(\vec{r})
\label{Rashba}
\end{equation}
Following  \cite{Peierls} we use the explicit form of the deformation field introduced by the dislocations $u(y)=\frac{B^{(2)}}{\pi}\tan^{-1}(\frac{y}{\widetilde{W}}))$.
According  to the elasticity theory,   the  Peierls  edge dislocation  \cite{kosevich}  is determined by the    half width  $\widetilde{W}$   function and  Burger vector $B^{(2)}$: 
$\rho(y)dy=B^{(2)}$, $\rho(y)=-\frac{du}{dy}$,\hspace{0.2 in}
$u(y)\propto M(y)=\frac{B^{(2)}}{\pi}\tan^{-1}(\frac{y}{\widetilde{W}})$.
Replacing the coordinates $[x,y]$ by  the deformed one, $[X(\vec{r}),Y(\vec{r})]$ we introduce  the deformed matrix elements   $t^{'}\neq t^{'}_{0}$ :  $t^{'}=t^{'}_{0}+(\frac{\partial t^{'}_{0}}{\partial_{ u(y)}})u(y)$. As a result we find: 
$-2t'+M=(-2t'_{0}+M)-2(\frac{\partial t^{'}_{0}}{\partial_{ u(y)}})u(y)\equiv 0 -\hbar v_{F} \kappa u(y)$   
where   $\hbar v_{F}\equiv 2t_{0}$ , $\kappa= \frac{1}{t_{0}}(\frac{\partial t^{'}_{0}}{\partial_{ u(y)}})$.
As a result we obtain  the Peierls Hamiltonian which is  similar to the domain wall model introduced by   \cite{Jackiw}.
\begin{eqnarray}
&&\epsilon \psi(x,y)=h^{Peierls}(x,y)\psi(x,y)\nonumber\\&&
h^{Peierls}(x,y)=\hbar v_{F}[-i\sigma^{x}\partial_{y} +i\sigma^{y}\partial_{x}- \sigma^{z}\kappa M(y)]\nonumber\\&&
\sigma^{y} \psi_{\lambda}(x,y)=\lambda \psi_{\lambda}(x,y);\hspace{0.1 in} \lambda=\pm 1;\hspace{0.2 in} \nonumber\\&&
\end{eqnarray}
The eigenfunction  for the Peierls model obeys periodic boundary conditions in the $x$ direction  $\psi(x+L,y)=\psi(x,y)$. Therefore  the eigen- spinor is given by :
$\psi(x,y)=e^{ip x}\Phi(y),\hspace{0.2 in} p=\frac{2\pi}{L}m, \hspace{0.2 in} m=0,\pm 1,\pm 2,...$.
The eigenvalue   $\epsilon$ is  replaced by $E$ where $E \equiv\epsilon-\hbar v_{F} p$.
We observe that if $\psi(x,y)$ is an eigenfunction with the eigenvalue $E$, the eigenfunction  $\sigma^{y}\psi(x,y)$ corresponds to the  eigenvalue  $-E$. $E=0$ corresponds to   the eigenvalue $\epsilon= \hbar v_{F} p$  with the  zero mode wave function  $\psi_{0}(x,y;p)$ given by :  
\begin{eqnarray}
&&\psi_{0}(x,y;p)=  e^{ip x}(\theta[y]  e^{-\frac{\kappa B^{(2)}}{\pi}  \int_{0}^{y}\,dy' \tan^{-1}(\frac{y'}{\widetilde{W}})} \left(\begin{array}{cc}1\\i\end{array}\right)+\theta[-y]  e^{\frac{ \kappa B^{(2)}}{\pi}  \int_{0}^{y}\,dy'\tan^{-1}(\frac{y'}{\widetilde{W}})}\left(\begin{array}{cc}1\\-i\end{array}\right))\Phi(0)\nonumber\\&&
\end{eqnarray} 
$\theta[y]$  represents the step function which is one for $y>0$ and zero otherwise.
As a result, we obtain  for $B^{(2)}>0$   only a single  zero mode  with  $\lambda=1$  for $y<0$   and $\lambda=-1$  for $y>0$.
The  eigenstate given in eq.$(6)$,  $|U_{0}(p)> $  and  $|U_{0}(-p)>$  are related by the  time reversal symmetry  $T |U_{0}(p)>\propto |U_{0}(-p)>$. From  the equations  $T^2=-1$ and   $T |U_{0}(p)>\propto |U_{0}(-p)>$  we conclude that the zero mode state is stable against   backscattering along the $x$ direction. ( The states  $|U_{0}(p)> $  and  $|U_{0}(-p)>$  are defined according to their    wave functions  $<x,y|U_{0}(p)>\equiv \psi_{0}(x,y;p)$  , $<x,y|U_{0}(-p)>\equiv \psi_{0}(x,y;-p)$.) 

\vspace{0.2 in}

\textbf{B-The edge dislocation with  with a non zero torsion $T^{(2)}_{\mu,\nu}$}

\vspace{0.2 in}

For the remaining part, we will limit our discussion to the microscopic formulation of dislocations.
When an edge  dislocation  is introduced into the crystal,   the lattice coordinates  $ \vec{r}=(x,y)$ are modified, $ \vec{r}\rightarrow \vec{R}=\vec{r}+\vec{u}\equiv[X(\vec{r}),Y(\vec{r})]$ where $\vec{u}(\vec{r})$ is the local lattice deformation with the core at  the dislocation centered at $\vec{r}=(0,0)$.  The elastic field contour integral around the core  $\vec{r}=(0,0)$  is determined by the Burger vector  $B^{(a)}$, $a=1,2$: $\displaystyle\oint dx^{\mu} \partial_{\mu}u^{(a)}(\vec{r})=- B^{(a)}$. 
For an edge dislocation in the $x$ direction   the $Burger$ vector $B^{(2)}$  is in the $y$ direction . The value of the burger vector   $B^{(2)}$  is given by   the   shortest translation  lattice  vector   in the $y$ direction.  (For  the  $TI$  $Bi_{2}Se_{3}$ the length of the vector  $B^{(2)}$ is $5$ times  the inter atomic  distance ).
For the two dimensional surface we use the notation $ x^{\mu}$ ,$\mu=x,y$ for the \textbf{fixed  Cartesian  coordinate } and $X^{a}$ ,$a=1,2$ to describe the \textbf{ reference frame }, the media with dislocations. In the media with dislocations we introduced a set of vectors $e_{a}$ which are orthogonal to each other $(e_{b},e_{a})\equiv <e^{b}|e_{a}>=\delta^{b}_{a}$. The unit vector $e_{a}$ can be represented in terms of the Cartesian fixed frame space with the coordinate basis $\partial_{\mu}$ ,$ \mu=x,y$. The basis in the media frame  can be expanded in terms of the  fixed Cartesian frame $\partial_{\mu}$; we have : $e_{a}=e^{\mu}_{a}\partial_{\mu}$  (for  the particular case where  vectors $e_{a}$ are given by  $e_{a}=\partial_{a}$  the transformation between the two basis is  $e^{\mu}_{a}=\delta^{\mu}_{a}$).
Any vector $\vec{X} $ can be represented in terms of the unit vectors $e_{a}$ (in the dislocation space)  and $\partial_{\mu}$ in the Cartesian  fixed coordinates  space , $\vec{X}=X^{a}e_{a}=X^{\mu}\partial_{\mu} $. The dual vector  $e^{a}$ is a $one$ $form$ and can be expanded in terms of the one forms  $dx^{\mu}$. We have $e^{a}=e^{a}_{\mu}dx$,  where $ e^{a}_{\mu}$ represents the matrix transformation $e^{a}\equiv (\partial_{\mu} X^{a})dx^{\mu}$. 
The scalar product of the components   $e^{a}_{\mu}e^{a}_{\nu}=g_{\mu,\nu}$, $e^{\nu}_{a}e^{\nu}_{b}=\delta_{a,b}$  defines the metric tensors, $g_{\mu,\nu}$  (in the Cartesian  frame ) and  $\delta_{a,b}$ in the dislocation frame.  
We will compute  the  matrix elements  fields $ e^{a}_{\mu}$ for our problem  :
\begin{equation}
e^{a}_{\mu}=\partial_{\mu}X^{a}(\vec{r}); \hspace{0.1 in} a=1,2 ; \hspace{0.1 in}  \mu=x,y
\label{ea}
\end{equation}
Following  \cite{Nelson} we can express the Burger vector  in terms of the the partial derivatives  with respect the coordinates $a=1,2$ in the dislocation frame and $\mu=x,y$ for the fixed Cartesian frame :
\begin{equation} 
 \partial_{x}e^{2}_{y}-\partial_{y}e^{2}_{x} = B^{(2)}\delta^{2}(\vec{r})
\label{stokes}
\end{equation} 
Using Stokes theorem, we  replace the line integral $\displaystyle\oint dx^{\mu}e^{2}_{\mu}(\vec{r})$  by the surface integral   $\int\int dx^{\mu}
 dx^{\nu}[\partial_{x}e^{2}_{y}-\partial_{y}e^{2}_{x}]  $. For a system with zero $curvature$   and non zero $torsion$  $T^{(2)}_{\mu,\nu}$ we find that  the  surface torsion tensor integral $\int\int dx^{\mu}
 dx^{\nu}T^{(2)}_{\mu,\nu}$  is equal to  $\int\int dx^{\mu}
 dx^{\nu}[\partial_{x}e^{2}_{y}-\partial_{y}e^{2}_{x}]  $,  and therefore both   integrals are equal  to  the Burger vector.
\begin{eqnarray}
&&\displaystyle\oint dx^{\mu}e^{2}_{\mu}(\vec{r})= \int\int dx^{\mu} dx^{\nu}[\partial_{\mu}e^{2}_{\nu}-\partial_{\nu}e^{2}_{\mu}]= B^{(2)};\nonumber\\&&
\int\int dx^{\mu}
 dx^{\nu}T^{(2)}_{\mu,\nu}=\int\int dx^{\mu} dx^{\nu}[\partial_{\mu}e^{2}_{\nu}-\partial_{\nu}e^{2}_{\mu}]= B^{(2)};\nonumber\\&&
\partial_{x}e^{2}_{y}-\partial_{y}e^{2}_{x}=B^{(2)}\delta^{2}(\vec{r})\nonumber\\&&
\end{eqnarray}
where $dx^{\mu}dx^{\nu}$ represents the surface element.
The  tangent components  $e^{a}_{\mu}$  can be expressed in terms  of the Burger  vector  density $ B^{(2)}\delta^{2}(\vec{r})$  \cite{Kleinert} :
\begin{eqnarray}
&&e^{2}_{x}=(\frac{B^{(2)}}{2\pi})\frac{y}{(x^2+y^2)};\hspace{0.3 in} e^{2}_{y}=1-(\frac{B^{(2)}}{2\pi})\frac{x}{(x^2+y^2)} \nonumber\\&&e^{1}_{x}=1 ;\hspace{0.4 in}e^{1}_{y}=0
\end{eqnarray} 
Using the tangent components, we obtain the metric tensor $g_{\mu,\nu}$. 
\begin{equation}
e^{a}_{\mu}e^{a}_{\nu}\equiv e^{1}_{\mu}e^{1}_{\nu}+ e^{2}_{\mu}e^{2}_{\nu}=g_{\mu,\nu}(\vec{r});\hspace{0.1 in}
e^{a}_{\mu}e^{b}_{\mu}\equiv e^{a}_{x}e^{b}_{x}+ e^{a}_{y}e^{b}_{y}=\delta_{a,b}
\label{ec}
\end{equation} 
The inverse of the metric tensor  $g_{\mu,\nu}(\vec{r})$ is the tensor  $g^{\nu,\mu}(\vec{r})$
defined trough the equation $g_{\mu,\tau}(\vec{r})g^{\tau,\nu}(\vec{r})=\delta_{\mu}^{\nu}$.
Using the tangent vectors,  we find   $to$ $first$ $order$ in the Burger vector the metric tensor $g_{\mu,\nu}$ and the  Jacobian transformation $\sqrt{G}$:
\begin{equation}
g_{x,x}=1;\hspace {0.1 in} g_{x,y}=\frac{B^{(2)}}{2\pi}\frac{y}{x^{2}+y^{2}};\hspace{0.1 in}  g_{y,y}=1-\frac{B^{(2)}}{2\pi}\frac{y}{x^{2}+y^{2}}; \hspace{0.1 in} g_{y,x}=0; \hspace{0.1 in}
G=det[g_{\mu,\nu}]=1-\frac{B^{(2)}}{2\pi}\frac{y}{x^{2}+y^{2}}
\label{metric}
\end{equation}
The inverse tensor is given by:$g^{x,x}\approx 1$, $g^{x,y}=g^{y,x}=-\frac{B^{(2)}}{2\pi}\frac{y}{x^{2}+y^{2}}$, $g^{y,y}=1+\frac{B^{(2)}}{\pi}\frac{x}{x^{2}+y^{2}}$.
Using the inverse tensor $ g^{\mu,\nu}$ we obtain the inverse matrix $e^{\mu}_{a} $ which is given by:
\begin{equation}
e^{\mu}_{a}=e_{a,\nu}g^{\nu,\mu}=(\delta_{a,b}e^{b}_{\nu})g^{\nu,\mu}=e^{a}_{\nu}g^{\nu,\mu}
\label{trans}
\end{equation}
In figure 1 we show the coordinate transformation  $\vec{r}=(x,y)\rightarrow  [X(\vec{r})=x,Y(\vec{r})=y+\frac{B^{(2)}}{2\pi}\tan^{-1}(\frac{y}{x})]$ with the core of the  dislocation centered at $\vec{r}=(0,0)$.

 Next we consider the effect of the dislocation on the $TI$ Hamiltonian.
Using the  components $e^{\mu}_{a} $  we compute the  the transformed Pauli matrices.
The Hamiltonian in the absence of the edge  dislocation is given by $h^{T.I.}=i \gamma^{a}\partial_{a}\equiv \sum_{a=1,2} i\gamma^{a}\partial_{a}$ where the Pauli matrices  are given by  $\gamma^{1}=-\sigma^{2}$ , $\gamma_{2}=\sigma_{1}$  and $\gamma^{3}=\sigma^{3}$. (We will use the convention that when an index appears twice we  perform a summation  over this index.)  In the presence of the   edge dislocation, the term $\gamma^{a}\partial_{a}$  must   be expressed in terms of the Cartesian   fixed coordinates  $\mu=x,y,z$. As a result, the spinor $\Psi(\vec{r})$    transforms accordingly to the $SU(2)$ transformation . If $\widetilde{\Psi}(\vec{R})$ is the spinor for the deformed lattice, it can be related with the help of an $SU(2)$ transformation to the spinor  $\Psi(\vec{r})$ in the  undeformed lattice:  $\widetilde{\Psi}(X,Y)=e^{-i\frac{\delta\varphi(x,y)}{2}\sigma^{3}}\Psi(x,y)$ .  Where   $\delta\varphi(x,y)$ is the rotation angle between the two  set of coordinates:
$\delta\varphi(x,y)= tan^{-1}(\frac{Y}{X})-tan^{-1}(\frac{y}{x})$. Using the relation between the coordinates (see figure 1) $X=x$, and $Y=y+\frac{B^{(2)}}{2\pi}tan^{-1}(\frac{y}{x})$ with the singularity at $x=y=0$  gives us that the derivative of the  phase which is a delta function, $\partial_{x}\delta\varphi(x,y)=-\partial_{y}\delta\varphi(x,y)\propto \delta^{2}(x,y)$. Combining the  transformation of the derivative with the $SO(2)$ rotation in the plane,  we obtain  the form of the chiral Dirac equation in the Cartesian space  (see Appendix A) given in terms of the  $spin$ $connection$  $\omega_{\mu}^{1, 2}$   \cite{Nakahara}:
\begin{equation}
i\gamma_{a}\partial_{a}\widetilde{\Psi}(\vec{R})= i\delta_{a,b}\gamma^{b}\partial_{a}\widetilde{\Psi}(\vec{R})=i\gamma^{a}e^{\mu}_{a}[\partial_{\mu}+\frac{1}{4}[\gamma^{b},\gamma^{c}]\omega_{\mu}^{b c}]\Psi(\vec{r})
\label{trans}
\end{equation}
The Hamiltonian $h^{T.I.}\rightarrow h^{edge}$  is transformed to the dislocation edge Hamiltonian  with the explicit form given by:
\begin{eqnarray}
&&h^{edge}=i\sigma^{1}\partial_{2}-i\sigma^{2}\partial_{1}=
i\sigma^{1}e^{\mu}_{2}[\partial_{\mu}+\frac{1}{8}[\sigma^{1},\sigma^{1}]\omega ^{1,2}_{\mu}]-i\sigma^{2}e^{\mu}_{1}[\partial_{\mu}+\frac{1}{8}[\sigma^{1},\sigma^{1}]\omega ^{1,2}_{\mu}]\nonumber\\&&=
i(\sigma^{1}e^{\mu}_{2}-\sigma^{2}e^{\mu}_{1})
(\partial_{\mu}+\frac{1}{8}[\sigma^{1},\sigma^{2}]\omega ^{1,2}_{\mu})
\nonumber\\&&
\end{eqnarray}
To first  order in the Burger vector we find : $\omega_{x}^{1 2}=- \omega_{x}^{2 1 }=0$ and $- \omega_{y}^{2 1 }=\omega_{y}^{1, 2}= -\frac{B^{(2)}}{2}\delta^{2}(\vec{r}) $, see eqs. $(69-71)$  in Appendix A.
\begin{equation}
h^{edge}\approx i\sigma^{1}(\partial_{y} -\frac{i}{2} \sigma^{3}B^{(2)}\delta^{2}(\vec{r})) -i\sigma^{2}\partial_{x}
\label{Burger}
\end{equation}
In the second quantized form  the chiral Dirac Hamiltonian in the presence of an edge dislocations  is given by :
\begin{eqnarray}
&&H^{edge}\approx \int\,d^{2}r \sqrt{G}\Psi^{\dagger}(\vec{r})[h^{edge}-\mu]\Psi(\vec{r})\nonumber\\&&\equiv \hbar v_{F}\int\,d^{2}r \sqrt{G}\Psi^{\dagger}(\vec{r})[i\sigma^{1}(\partial_{y} -\frac{i}{2} \sigma^{3}B^{(2)}\delta^{2}(\vec{r}))-i\sigma^{2}\partial_{x} -\mu ]\Psi(\vec{r})\nonumber\\&&
\end{eqnarray}
$h^{edge}$
is the Hamiltonian in the first quantized language, $\mu$ is the chemical potential and $\Psi(\vec{r})=[\Psi_{\uparrow}(\vec{r}),\Psi_{\downarrow}(\vec{r})]^{T}$ is the two component spinor field.

\vspace{0.2 in}

\textbf{IV- The solution for    the  metallic Surface in the presence of an   edge dislocation  with  the torsion  $T^{(2)}_{\mu,\nu}$}

\vspace{0.2 in}

\textbf{A-The   model in the momentum space}

\vspace{0.2 in}

We work in the  momentum representation \cite{Spinorbit}  where   the edge Hamiltonian  $h^{edge}$ takes the form:
\begin{equation}
h^{edge}=\hbar v_{F}[( \vec{K}\times \vec{\sigma})\cdot\hat{z}]+ \frac{-i}{2}B^{(2)}\sigma^{2}\int\,\frac{d^{2}q}{(2\pi)^2}e^{i[q_{x}(i\partial_{k_{x}})+q_{y}(i\partial_{k_{y}})]}
\label{edgp}
\end{equation}  
The  Hamiltonian  is \textbf{time reversal invariant}  but it is \textbf{ not invariant under the  planar parity symmetry}. 
The eigenstate $|u(\vec{K})>$ obeys the secular equation:   $[\epsilon-\epsilon(\vec{K})+i\frac{B^{(2)}}{2}\sigma^{2}]|W(\epsilon)>=0$ where 
$|W(\epsilon)>\equiv\int\frac{d^{2}q}{(2\pi)^2}|u(\vec{K}-\vec{q})>$.  
The  the presence of the term  $\frac{-i}{2}B^{(2)}\sigma^{2}$ generates   unstable solutions . A stable  solution can be obtained if the Pauli matrix $\sigma^{2}$ annihilates the eigenstate. Using  the chirality operator  we observe that states propagating in the $x$ direction are eigenstates of  $\sigma^{2}$  (states polarized in the $y$ direction).
$\sigma^{2} |\vec{K}=k_{x}>\otimes[1,i e^{i\chi(k_{x},k_{y}=0)}]^{T}= sgn(k_{x})|\vec{K}=k_{x}>\otimes[1,i e^{i\chi(k_{x},k_{y}=0)}]^{T}$.
The eigenstates of the edge Hamiltonian  which satisfy the chirality operator  must have  the form :  $|u(k_{x},y)>=e^{i k_{x} x}[|u_{\uparrow}(y)>,|u_{\downarrow}(y)>]^{T}$.
In the next sections we will construct the explicit eigenstates for this Hamiltonian.

\vspace{0.2 in}

\textbf{B-Identification of the physical contours for the  edge  Hamiltonian $h^{edge}$}

\vspace{0.2 in}

In order to identify the solutions, we will use the complex representation.
The coordinates in   the   complex representation are given by,
$z=\frac{1}{2}(x+iy)$,\hspace{0.1 in} $\overline{z}=\frac{1}{2}(x-iy)$, \hspace{0.1 in}
$\partial_{z}=\partial_{x}-i\partial_{y}$, \hspace{0.1 in} $\partial_{\overline{z}}=\partial_{x}+i\partial_{y}$. In this representation    the two dimensional  delta function  $\delta^{2}(\vec{r})$ is given by  $\delta^{2}(\vec{r})\equiv\frac{1}{\pi} 
\partial_{z}(\frac{1}{ \overline{z}})=\frac{1}{\pi} 
\partial_{\overline{z}}(\frac{1}{z})$ \cite{Conformal,Nair}.
We will  use the edge Hamiltonian  $h^{edge}   $
and will compute the eigenfunctions  $u_{\epsilon}(z,\overline{z})=[U_{\epsilon\uparrow}(z,\overline{z}),U_{\epsilon\downarrow}(z,\overline{z})]^{T}$ and $v_{\epsilon}(z,\overline{z})=V_{\epsilon\uparrow}(z,\overline{z}),V_{\epsilon\downarrow}(z,\overline{z})]^{T}$.
The eigenvalue equation is given by:
\begin{eqnarray}
&&\epsilon U_{\epsilon\uparrow}(z,\overline{z})=-[\partial_{z}+(\frac{B^{(2)}}{ \sqrt{2}\pi})\partial_{z}(\frac{1}{ \overline{z}})] U_{\epsilon\downarrow}(z,\overline{z})\nonumber\\&&
 \epsilon U_{\epsilon \downarrow}(z,\overline{z})=[\partial_{\overline{z}}+ (\frac{ B^{(2)}}{\sqrt{2}\pi})\partial_{\overline{z}}(\frac{1}{z})] U_{\epsilon\uparrow}(z,\overline{z})\nonumber\\&&
\end{eqnarray} 
The eigenfunctions  $u_{\epsilon}(z,\overline{z})$ and $v_{\epsilon}(z,\overline{z})$ can be written with the help of   a singular  matrix $ M(z,\overline{z})$ \cite{Ezawa} :
\begin{equation*}
u_{\epsilon}(z,\overline{z})=M(z,\overline{z})\hat{F}_{\epsilon}(z,\overline{z})\equiv\left[\begin{array}{rrr}
e^{-\frac{B^{(2)}}{2\pi}(\frac{1}{ z})}& 0 \\
0 & e^{-\frac{B^{(2)}}{2\pi}(\frac{1}{ \overline{z}})}\\
\end{array}\right]
\left(\begin{array}{cc}F_{\epsilon\uparrow}(z,\overline{z})\\F_{\epsilon\downarrow}(z,\overline{z})\end{array}\right)
\end{equation*}
 ($F_{\epsilon}(z,\overline{z})$ and $F_{-\epsilon}(z,\overline{z})$  are the   transformed eigenfunctions for $\epsilon>0$ and   $\epsilon <0$  respectively .) In terms of the transformed spinors 
the eigenvalue equation $h^{edge}(z,\overline{z})u_{\epsilon}(z,\overline{z})=\epsilon u_{\epsilon}(z,\overline{z})$ and $F_{\epsilon\downarrow}(z,\overline{z})$ becomes: 
\begin{equation*}
\epsilon\left(\begin{array}{cc}F_{\epsilon\uparrow}(z,\overline{z})\\F_{\epsilon\downarrow}(z,\overline{z})\end{array}\right)=\left[\begin{array}{rrr}
I(z,\overline{z})& 0 \\
0 & (I(z,\overline{z})^{*}\\
\end{array}\right]\left[\begin{array}{rrr}
-\partial_{z} & 0 \\
0 & \partial_{\overline{z}} \\
\end{array}\right]\left(\begin {array}{cc}F_{\epsilon\uparrow}(z\overline{z})\\F_{\epsilon\downarrow},(z,\overline{z})\end{array}\right)
\end{equation*}
where  $I(z,\overline{z})= e^{-\frac{B^{(2)}}{2\pi}(\frac{\overline{z}-z}{ z \overline{z}})}\equiv e^{2\frac{B^{(2)}}{\pi}(\frac{ iy}{x^2+y^2})}$ , $(I(z,\overline{z}))^{*}= e^{2\frac{B^{(2)}}{\pi}(\frac{-iy}{x^2+y^2})}$, $|I(z,\overline{z})|=1$. 
We search for zero modes  $\epsilon=0$ and find  :
\begin{equation}
\partial_{z}F_{\epsilon\downarrow}(z,\overline{z})=0\hspace{0.5 in}
\partial_{\overline{z}}F_{\epsilon\uparrow}(z,\overline{z})=0
\label{zero}
\end{equation}
The solutions are given by the holomorphic   representation 
$F_{\epsilon=0\uparrow}(z,\overline{z})=f_{\uparrow}(z)$  and    the anti-holomorphic function   $F_{\epsilon=0\downarrow}(z,\overline{z})=f_{\downarrow}(\overline{z})$.
The  zero mode eigenfunctions are  given by :
\begin{equation}
u_{\epsilon=0,\uparrow}(z)=e^{-\frac{ B^{(2)}}{2\pi}(\frac{1}{z})} f_{\uparrow}(z),\hspace{0.5 in}
u_{\epsilon=0,\downarrow}(\overline{z})=e^{-\frac{ B^{(2)}}{2\pi}(\frac{1}{\overline{z}})}f_{\downarrow}(\overline{z})
\label{eigen}
\end{equation}
Due to the presence of the essential singularity at $z=0$ it is not possible to find  analytic functions  $f_{\uparrow}(z)$  and   $f_{\downarrow}(\overline{z})$   which  vanish fast enough around   $z=0$ such that $\int d^{2}z (u_{\epsilon=0,\lambda}(z))^{*} u_{\epsilon=0,\lambda}(z)<\infty $. Therefore, we conclude that  \textbf{zero mode solution} does not exists.    
The only way to remedy the problem is to allow for states with finite energy. 

 In the next step  we  look for finite energy states.
We perform a coordinate transformation :
\begin{equation}
z\rightarrow W[z,\overline{z}];\hspace{0.5 in}  \overline {z}\rightarrow\overline{W}[z,\overline{z}]
\label{transform}
\end{equation}
We  demand that the transformation is conformal and preserve the orientation. This restricts the transformations to   holomorphic and anti-holomorphic functions \cite{Conformal}. This means that  we  have the conditions    $\partial_{\overline{z}} W[z,\overline{z}]=0 $ and $\partial_{z} \overline{W}[z,\overline{z}]=0 $. As a result  we obtain   $W[z,\overline{z}]=W[z]$ and  $\overline{W}[z,\overline{z}]=\overline{W}[\overline{z}]$, which obey  the eigenvalue equations: 
\begin{eqnarray}
&&\epsilon F_{\epsilon\uparrow}(W,\overline{W})=-\partial_{W}F_{\epsilon\downarrow}(W,\overline{W})\nonumber\\&&
\epsilon F_{\epsilon\downarrow}(W,\overline{W})= \partial_{\overline {W}}F_{\epsilon\uparrow}(W,\overline{W})\nonumber\\&&
\end{eqnarray}
This implies the conditions $\frac{dW[z]}{dz}=(I(z,\overline{z}))^{*}$ and  $\frac{d\overline{W}[\overline{z}]}{dz}=I(z,\overline{z})$. Since  $I(z,\overline{z})$  is neither holomorphic or anti-holomorphic  and satisfy    $|I(z,\overline{z})|=1$, the only   solutions for $W[z]$ and $\overline{W}[\overline{z}]$ must obey      $I(z,\overline{z})=1$: 
\begin{equation}
I(z,\overline{z})\equiv e^{2\frac{B^{(2)}}{\pi}(\frac{ iy}{x^2+y^2})}=
e^{i 2\pi n};\hspace{0.2 in} n=0,\pm 1,\pm 2....
\label{solution}
\end{equation} 
For   $I(z,\overline{z})\neq 1$  one obtains solutions which are unstable .    The stable solutions  will be given by a  one parameter $s$   curve  ($s$ is the length of the curve)  $\vec{r}(s)\equiv[x(s),y(s)]$      which obey the equation  $I(z,\overline{z})=1$. 
 The curve $\vec{r}(s)$  allows  us to  define the $tangent$ $\vec{t}(s)$  and the  $normal$ vectors   $\vec{N}(s)$  \cite{Exner}. This allows us  to introduce   a two- dimensional region in the vicinity of the contour  of $\vec{r}(s)\rightarrow  \vec{R}(s,u)=\vec{r}(s)+u \vec{N}(s)$.

\vspace{0.2 in}

\textbf{V- The wave function}

\vspace{0.2 in}

\textbf{A-The  wave function   for  $n=0$}

\vspace{0.2 in}

 The condition $I(z,\overline{z})=e^{2\frac{B^{(2)}}{\pi}(\frac{ iy}{x^2+y^2})} =1$   for   $n=0$ is satisfied  for  $y=0$ and large  value of $y$  which obey    $2\frac{B^{(2)}}{\pi}(\frac{ y}{x^2+y^2})<<1$ .  The  values of $y$   which  satisfy   this conditions are restricted to   $I(z,\overline{z})=e^{2\frac{B^{(2)}}{\pi}(\frac{ iy}{x^2+y^2})}\approx 1$.
This  condition is satisfied  for values of  $y$  in the range:
\begin{equation} 
2\frac{B^{(2)}}{\pi}(\frac{ y}{x^2+y^2})\leq \eta <\frac{\pi}{4}<1
\label{eta}
\end{equation}
We introduce the radius   $R_{g}=\frac{B^{2}}{2\pi^2}$  and find that the condition   $I(z,\overline{z})\approx1$ give rise to the  equation for $y$. The solution   is given by  $ x^2+ (y\pm \frac{2\pi}{\eta}R_{g})^2 =( \frac{2\pi}{\eta}R_{g})^2$. 
Therefore,  for $|y|>|d|\geq(\frac{2\pi}{\eta}) 2R_{g}> 2R_{g}$ we have   $I\approx1$ which corresponds to   a free particle  eigenvalue equations.
\begin{eqnarray}
&&\epsilon F_{\epsilon\uparrow}(x,y)=e^{\frac{ B^{(2)}}{\pi}  \frac{i2y}{(x^2+y^2)}}[-\partial_{x} +i\partial_{y}]F_{\epsilon\downarrow}(x,y)\nonumber\\&&
\approx  [-\partial_{x} +i\partial_{y}]F_{\epsilon\downarrow}(x,y);\nonumber\\&& \epsilon F_{\epsilon\downarrow}(x,y)=  e^{\frac{ B^{(2)}}{\pi}  \frac{-i2y}{(x^2+y^2)}}[\partial_{x}+i\partial_{y}]F_{\epsilon\uparrow}(x,y)\nonumber\\&&\approx[\partial_{x}+i\partial_{y}]F_{\epsilon\uparrow}(x,y)\nonumber\\&&
\end{eqnarray}
For $|y|>d$  the eigenfunctions are given by:
$U_{\epsilon,\uparrow}(x,y)=e^{-\frac{B^{(2)}}{2\pi}(\frac{1}{x+iy})}F_{\epsilon,\uparrow}(x,y)$, $U_{\epsilon,\downarrow}(x,y)=e^{-\frac{B^{(2)}}{2\pi}(\frac{1}{x-iy})}F_{\epsilon,\downarrow}(x,y)$ where $F_{\epsilon\uparrow}(x,y)$ and $F_{\epsilon\downarrow}(x,y)$ are  the eigenfunctions of  equation (26).   The  envelope functions  $e^{-\frac{ B^{(2)}}{2\pi}(\frac{1}{x+iy})}$, $e^{-\frac{ B^{(2)}}{2\pi}(\frac{1}{x-iy})}$  which  multiply  the wave functions   $F_{\epsilon\uparrow}(x,y)$ , $F_{\epsilon\downarrow}(x,y)$  impose  vanishing  boundary  conditions for the  eigenfunctions $U_{\epsilon,\downarrow}(x,y)$ and $U_{\epsilon,\uparrow}(x,y)$ at  $y\rightarrow\pm \infty$ .   therefore, we   demand that the  eigenfunctions $U_{\epsilon,\uparrow}(x,y)$, $U_{\epsilon,\downarrow}(x,y)$ should   vanish  at the boundaries  $y=\pm\frac{L}{2}$.
Since the multiplicative envelope functions for opposite spins is  complex conjugate to each other we have to make the choice that one of the spin components vanishes at one side and the other component at the opposite side. Two possible choices   can be made: 

$U_{\epsilon,\uparrow}(x,y=\frac{L}{2})\equiv e^{-\frac{ B^{(2)}}{2\pi}(\frac{1}{x+i\frac{L}{2}})}F_{\epsilon\uparrow}(x,\frac{L}{2})=  U_{\epsilon,\downarrow}(x,y=-\frac{L}{2})\equiv e^{-\frac{ B^{(2)}}{2\pi}(\frac{1}{x-i(-\frac{L}{2})})}F_{\downarrow}(x,-\frac{L}{2})=0$ 
 
\textbf{or}

$U_{\epsilon,\uparrow}(x,y=-\frac{L}{2})\equiv e^{-\frac{ B^{(2)}}{2\pi}(\frac{1}{x+i(-\frac{L}{2})})}F_{\epsilon\uparrow}(x,-\frac{L}{2})=  U_{\epsilon,\downarrow}(x,y=\frac{L}{2})\equiv e^{-\frac{ B^{(2)}}{2\pi}(\frac{1}{x-i\frac{L}{2}})}F_{\downarrow}(x,\frac{L}{2})=0$

Making the first choice, (both choices give the same eigenvalues and eigenfunction)  we compute the eigenfunctions $F_{\epsilon\uparrow}(x,y)$ and $F_{\epsilon\downarrow}(x,y)$ for $|y|>d$ using the boundary conditions :  
\begin{equation}  
F_{\epsilon\uparrow}(x,y=\frac{L}{2})=0;\hspace{0.4 in}
F_{\epsilon\downarrow}(x,y=-\frac{L}{2})=0 
\label{equabound}
\end{equation} 
Due to the fact that the solutions are restricted to $|y|>d$ no  conditions need to be imposed   at $x=y=0$.  
In the present case we consider a situation with a single dislocation. This is justified for a dilute concentration of  dislocations  typically separated by  a distance $l\approx 10^{-6} m$.  ( In principle we need at least two dislocations in order to satisfy the condition that the  sum of the Burger vectors is zero.) 
The eigenvalues are given by $\epsilon=\pm\hbar v_{F}\sqrt{p^2+q^2}$.  The value of $p$ is determined by the periodic boundary condition in the $x$ direction       $p(m)=\frac{2\pi}{L}m\equiv \frac{2\pi}{N a}m$, $m=0,1,...,(N-2),(N-1)$ and $a$ is the lattice constant  $a\approx \frac{2\pi}{\Lambda}$. The value of $q$ 
will be obtained  from the  vanishing boundary conditions  at $y=\pm\frac{L}{2}$.
The eigenfunctions $F_{\epsilon,\sigma}(x,y)$ will  be obtained using the  linear combination  of the  spinors  introduced in chapter $II$. In the Cartesian representation we can build four spinors  $\Gamma_{p,q}(x,y)$, $\Gamma_{p,-q}(x,y)$,$\Gamma_{-p,q}(x,y)$,$\Gamma_{-p,-q}(x,y)$  which are eigenstates of the chirality operator and are given by:
\begin{equation*}
\Gamma_{p,q}(x,y)=e^{i p x}e^{iq y} \left(\begin{array}{cc} 1\\ i e^{i\chi(p,q)}\end{array}
\right)
\end{equation*}
\begin{equation*}
\Gamma_{p,-q}(x,y)=e^{i p x}e^{-iq y}\left(\begin{array}{cc} 1\\ i e^{-i\chi(p,q)}\end{array}
\right)
\end{equation*}
\begin{equation*}
\Gamma_{-p,q}(x,y)=e^{i p x}e^{iq y}\left(\begin{array}{cc} 1\\ -i e^{-i\chi(p,q)}\end{array}
\right)
\end{equation*}
\begin{equation}
\Gamma_{-p,-q}(x,y)=e^{-i p x}e^{-iq y}\left(\begin{array}{cc} 1\\ -i e^{i\chi(p,q)}\end{array}
\right)
\label{spinors}
\end{equation}
where $tan[\chi(p,q)]=\frac{q}{p}$.

 The Hamiltonian $h^{edge}$ is \textbf{not invariant} under the symmetry  operation  
$P_{x}: x\rightarrow -x,\sigma^{1}\rightarrow \sigma^{1},\sigma^{2}\rightarrow -\sigma^{2}$;
$[h^{edge},P_{x}]\neq0$ therefore;
we  need to  construct two independent   eigenfunctions $F^{(n=0,R)}_{p>0,q}(x,y)$ for $p>0$  and $ F^{(n=0,L)}_{-p>0,q}(x,y)$  $p<0$.
\begin{eqnarray}
&&F^{(n=0,R)}_{p>0,q}(x,y)=A(q)\Gamma_{p,q}(x,y)+B(q)\Gamma_{p,-q}(x,y)\nonumber\\&&
F^{(n=0,L)}_{-p>0,q}(x,y)=C(q)\Gamma_{-p,q}(x,y)+D(q)\Gamma_{-p,-q}(x,y)\nonumber\\&&
\end{eqnarray}
Employing  the boundary conditions given in equation $(27)$ we obtain the amplitudes $\frac{D(q)}{C(q)}$ ,  $\frac{B(q)}{A(q)}$  and the conditions for the momenta $q$.
For the pair $ \Gamma_{p,q}(x,y)$ , $\Gamma_{p,-q}(x,y)$ $p>0$ we obtain :
\begin{eqnarray}
&&F^{(n=0,R)}_{\epsilon(p>0,q_{+}),\uparrow}(x,y)=e^{ipx}e^{\frac{i}{2}\chi(p,q_{+})}[e^{i(q_{+}y-\frac{1}{2}\chi(p,q_{+}))}+(-1)^{k+1}e^{-i(q_{+}y-\frac{1}{2}\chi(p,q_{+}))}]; |y|>d\nonumber\\&&
F^{(n=0,R)}_{\epsilon(p>0,q_{+}),\downarrow}(x,y)=ie^{ipx}e^{\frac{i}{2}\chi(p,q_{+})}[e^{i(q_{+}y+\frac{1}{2}\chi(p,q_{+}))}+(-1)^{k+1}e^{-i(q_{+}y+\frac{1}{2}\chi(p,q_{+}))}]; |y|>d\nonumber\\&&
q\equiv q_{+}=\frac{\pi}{L}k+\frac{1}{L}\tan^{-1}(\frac{q_{+}}{p});k=1,2,3...;\tan[\chi(p,q_{+})]=(\frac{q_{+}}{p})\nonumber\\&&
\epsilon(p,q_{+})=\pm\hbar v_{F}\sqrt{(\frac{2\pi}{L}m)^2+q_{+}^2}\nonumber\\&&
\end{eqnarray}
Similarly, for the second  pair $ \Gamma_{-p,q}(x,y)$,$\Gamma_{-p,-q}(x,y)$, $p>0$  we obtain:
\begin{eqnarray}
&&F^{(n=0,L)}_{\epsilon(-p>0,q_{-}),\uparrow}(x,y)=e^{-ipx}e^{-\frac{i}{2}\chi(p,q_{-})}[e^{i(q_{-}y+\frac{1}{2}\chi(p,q_{-}))}+(-1)^{k+1}e^{-i(q_{-}y+\frac{1}{2}\chi(p,q_{-}))}]; |y|>d\nonumber\\&&
F^{(n=0,L)}_{\epsilon(-p>0,q_{-}),\downarrow}(x,y)=-ie^{-ipx}e^{-\frac{i}{2}\chi(p,q_{-})}[e^{i(q_{-}y-\frac{1}{2}\chi(p,q_{-}))}+(-1)^{k+1}e^{-i(q_{-}y-\frac{1}{2}\chi(p,q_{-}))}]; |y|>d\nonumber\\&&
q\equiv q_{-}=\frac{\pi}{L}k-\frac{1}{L}\tan^{-1}(\frac{q_{-}}{p});k=1,2,3...;\tan[\chi(p,q_{-})]=(\frac{q_{-}}{p})\nonumber\\&&
\epsilon(-p,q_{-})=\pm\hbar v_{F}\sqrt{(\frac{2\pi}{L}m)^2+q_{-}^2}\nonumber\\&&
\end{eqnarray}
For the state  with zero momentum $p=0$ we find:
\begin{eqnarray}
&&F^{(n=0,0)}_{\epsilon(p=0,q),\uparrow}(x,y)=2 e^{\frac{-i\pi}{4}}\cos[qy+\frac{\pi}{4}]; |y|>d\nonumber\\&&
F^{(n=0,0)}_{\epsilon(p=0,q),\downarrow}(x,y)=i2e^{\frac{-i\pi}{4}}\cos[qy-\frac{\pi}{4}]; |y|>d\nonumber\\&&
q=\frac{\pi}{2L}+\frac{\pi}{L}k;k=0,1,2,3...\nonumber\\&&
\epsilon(p=0,q)=\pm\hbar v_{F}|q|\nonumber\\&&
\end{eqnarray}
The eigenfunctions for the dislocation problem will be given  for $|y|>d$  by:\hspace{0.15 in} $u^{(n=0,R)}_{\epsilon}(x,y)\equiv[U^{(n=0,R}_{\epsilon\uparrow}(x,y),U^{(n=0,R)}_{\epsilon\downarrow}(x,y)]^{T}$, $u^{(n=0,L)}_{\epsilon}(x,y)\equiv[U^{(n=0,L}_{\epsilon\uparrow}(x,y),U^{(n=0,L)}_{\epsilon\downarrow}(x,y)]^{T}$.
\begin{eqnarray}
&&U^{(n=0,R)}_{\uparrow}(x,y)\approx\frac{2 const.(B^{(2)})}{ G^{\frac{1}{4}}(x,y)L}
e^{-\frac{ B^{(2)}}{2\pi}(\frac{1}{x+iy})}F^{(n=0,R)}_{\epsilon(p>0,q_{+}),\uparrow}(x,y)\nonumber\\&&
U^{(n=0,R)}_{\downarrow}(x,y)\approx\frac{2 const.(B^{(2)})}{ G^{\frac{1}{4}}(x,y)L}
e^{-\frac{ B^{(2)}}{2\pi}(\frac{1}{x-iy})}F^{(n=0,R)}_{\epsilon(p>0,q_{+}),\downarrow}(x,y)\nonumber\\&&
U^{(n=0,L)}_{\uparrow}(x,y)\approx\frac{2 const.(B^{(2)})}{ G^{\frac{1}{4}}(x,y)L}
e^{-\frac{ B^{(2)}}{2\pi}(\frac{1}{x+iy})}F^{(n=0,L)}_{\epsilon(-p>0,q_{-}),\uparrow}(x,y)\nonumber\\&&
U^{(n=0,L)}_{\downarrow}(x,y)\approx\frac{2 const.(B^{(2)})}{ G^{\frac{1}{4}}(x,y)L}
e^{-\frac{ B^{(2)}}{2\pi}(\frac{1}{x-iy})}F^{(n=0,L)}_{\epsilon(-p>0,q_{-}),\downarrow}(x,y)\nonumber\\&&
U^{(n=0,0)}_{\uparrow}(x,y)\approx e^{-\frac{ B^{(2)}}{2\pi}(\frac{1}{x+iy})}F^{(n=0,0)}_{\epsilon(p=0,q),\uparrow}(x,y)\nonumber\\&&
U^{(n=0,0)}_{\downarrow}(x,y)\approx e^{-\frac{ B^{(2)}}{2\pi}(\frac{1}{x-iy})}F^{(n=0,0)}_{\epsilon(p=0,q),\downarrow}(x,y)\nonumber\\&&
\end{eqnarray}
where $G(x,y)=1-\frac{B^(2)}{2\pi}\frac{y}{\sqrt{2}(x^2+y^2)}$ is the Jacobian introduced by the edge dislocation.  The eigenstates are normalized and obey:$\int \,dx \int\,dy \sqrt{G(x,y)} (U^{(n=0,R)}_{\sigma}(x,y))^{*}  U^{(n=0,R)}_{\sigma'}(x,y)\approx \delta_{\sigma ,\sigma'}$, $\int \,dx \int\,dy \sqrt{G(x,y)} (U^{(n=0,L)}_{\sigma}(x,y))^{*}  U^{(n=0,L)}_{\sigma'}(x,y)\approx \delta_{\sigma ,\sigma'}$.
The  normalization  factor  $\frac{2 const.(B^{(2)} )}{L}\approx \frac{2  }{L}$,  has a weak  dependence  on the  Burger vector  $B^{(2)}$ . This dependence is a consequence of the Jacobian   $\sqrt{G}$ which affects  the normalization  constant (see  appendix $B$). 

(The  multiplicative factor  $e^{-\frac{ B^{(2)}}{2\pi}(\frac{1}{x \pm iy})}$  gives rise to a weak non-orthogonality  between the states.
 This    non-orthogonality  of the linear independent  eigenfunctions  can be corrected  with the help of  the Grahm-Shmidt method.)
 
 For the present case,  backscattering is allowed but it is much weaker in comparison to regular metals. This is seen as follows:
Time reversal is not violated; due to the parity violation, the  eigenstates  $u^{(n=0,R)}_{\epsilon}(x,y)$ ,$u^{(n=0,L)}_{\epsilon}(x,y)$ are \textbf{not} related by a time reversal symmetry ($T u^{(n=0,R)}_{\epsilon}(x,y)\neq  u^{(n=0,L)}_{\epsilon}(x,y)$) . As a result, the backscattering potential $V_{p,-p}$ is controlled by a finite matrix element   between states with different eigenvalues  $\epsilon(-p,q_{-})\neq  \epsilon(p,q_{+})$ (contrary to regular metals  where the impurity potential   $V_{p,-p}$ connects states with the same energy). In the present case $|\epsilon(-p,q_{-}) -   \epsilon(p,q_{+})|=\hbar v_{F}|[\sqrt{(\frac{2\pi}{L}m)^2+q_{-}^2}-\sqrt{(\frac{2\pi}{L}m)^2+q_{+}^2}]|\neq0$  the eigenvalues are not equal, therefore the finite matrix element controlled by the   backscattering potential  $V_{p,-p}$ gives rise only  to   a second order backscattering effect!

\vspace{0.2 in}
  
\textbf{B- The  circular  contours-the wave function for  $n\neq0$}

\vspace{0.2 in}

 The equation $I(z,\overline{z})=e^{2\frac{B^{(2)}}{\pi}(\frac{ iy(s)}{x^2(s)+y^2(s)})}  =e^{i 2\pi n}$  gives the set  of ring  contours  for $n=\pm1,\pm2,\pm3,...$ shown in figure 2.
The radius  $R_{g}$ for the fundamental contour($n=1$) is represented  in terms of the  Burger vector  $B^{(2)}$,   $R_{g}=\frac{B^{(2)}}{2\pi^2}$ and $R_{g}(n)=\frac{R_{g}}{|n|}$.
\begin{equation}
(x(s))^2+ (y(s)\pm R_{g}(n))^2 =(R_{g}(n))^2
\label{contour}
\end{equation}
The centers of the contours are given by :$[\bar{x},\bar{y}]=[0,R_{g}(n)]$  for $n\neq 0$. When $n>0$ the center of the contours has positive coordinates (upper contour) and for $n<0$  the center has negative coordinates (lower contour). 
Each contour is characterized by a circle with a radius $R_{g}(n)\equiv\frac{R_{g}}{|n|}$  centered at  $[\bar{x}=0,\bar{y}=R_{g}(n)]$ (see figure $2$). The contour is parametrized in terms of  the  arc length $0\leq s< 2\pi \frac{R_{g}}{|n|}$ which is  equivalent to  $0\leq\varphi<2\pi$ .
Each contour is parametrized  by  $\vec{r}(s)\equiv[x(s),R_{g}(n)+y(s)]$ where $x(s)= R_{g}(n) \cos[\frac{s}{R_{g}(n)}]\equiv\ R_{g}(n)\cos[\varphi]$ and $y(s)=  R_{g}(n) \sin[\frac{s}{R_{g}(n)}]\equiv\ R_{g}(n)\sin[\varphi]$. We will extend this curve to a two dimensional strip with the  coordinate $u$ in the normal direction:
For the curve  curve $\vec{r}(s)=[x(s),y(s)]$ we will use the tangent  $\vec{t}(s)$  and the normal vector  $\vec{N}(s)$  \cite{Exner}.  Therefore,   the  two dimensional region in the vicinity of the one parameter curve $\vec{r}(s)$ is replaced by   $\vec{r}(s)\rightarrow  \vec{R}(s,u)=\vec{r}(s)+u \vec{N}(s)$. 
\begin{eqnarray}
&&x(s,u)=R_{g}(n) \cos[\frac{s}{R_{g}(n)}]+u\cos[\frac{s}{R_{g}(n)}]\nonumber\\&&
y(s,u)=R_{g}(n) \sin[\frac{s}{R_{g}(n)}]+u\sin[\frac{s}{R_{g}(n)}]\nonumber\\&&
\end{eqnarray}
We will restrict the width $|u|$ such that $e^{i 2\pi n}e^{ \pm i\eta}\approx 1 $  where $\eta$ obeys  $\eta <\frac{\pi}{4}<1$ ,
  $|u|\leq \frac{R_{g}(n)}{1-\frac {\eta}{2\pi n}}-R_{g}(n)\approx R_{g}(n)(\frac {\eta}{2\pi n})< \frac{R_{g}(n)}{8n}$.  
In these new coordinates, the Dirac equation is approximated for  $|u|\leq R_{g}(n)(\frac {\eta}{2\pi n})=\frac{D(n)}{2}$ by :
\begin{eqnarray}
&&\epsilon F_{\epsilon \uparrow}(s,u)=-I(s,u)e^{-i\frac{s}{R_{g}(n)}}[\partial_{u}-\frac{i}{1+\frac{u}{R_{g}(n)}}\partial_{s}]F_{\epsilon\downarrow}(s,u)\approx -e^{-i\frac{s}{R_{g}(n)}}[\partial_{u}-i\partial_{s}]F_{\epsilon\downarrow}(s,u) \nonumber\\&&
\epsilon F_{\epsilon\downarrow}(s,u)=(I(s,u))^{*}e^{i\frac{s}{R_{g}(n)}}[\partial_{u}+\frac{i}{1+\frac{u}{R_{g}(n)}}\partial_{s}]F_{\epsilon\uparrow}(s,u)\approx e^{i\frac{s}{R_{g}(n)}}[\partial_{u}+i\partial_{s}]F_{\epsilon\uparrow}(s,u)\nonumber\\&&
\end{eqnarray}

\vspace{0.1 in}

\textbf{The solution  for the contour  $n\neq0$, $0\leq s <2\pi R_{g}(n)$; $|u|\leq \frac{D(n)}{2}$}

\vspace{0.1 in}

The periodicity in $s$ allows us to  represent the  eigenfunctions in  the form: $F_{\epsilon\uparrow}(s,u)=\sum_{j=-\infty}^{\infty}\sum_{q}e^{i j(\frac{s}{R_{g}(n)})}e^{iqu}F_{\epsilon\uparrow}(j,q)$
and $F_{\epsilon\downarrow}(s,u)=\sum_{j=-\infty}^{\infty}\sum_{q}e^{i (j+1)(\frac{s}{R_{g}(n)})}e^{iqu}F_{\epsilon\downarrow}(j,q)$. We find:
\begin{eqnarray}
&&\epsilon F_{\uparrow}(\epsilon;j,q)=(i q +\frac{j}{R_{g}(n)})F_{\downarrow}(\epsilon;j,q)\nonumber\\&&
\epsilon F_{\downarrow}(\epsilon;j,q)=(i q +\frac{j+1}{R_{g}(n)})F_{\uparrow}(\epsilon;j,q)\nonumber\\&&
\end{eqnarray}
The determinant of  the  two  equations  determines the relation between the eigenvalue $\epsilon$, the transverse momentum $Q(\epsilon)$ and the eigenfunctions $F_{\epsilon\downarrow}(j,q)$,$F_{\epsilon\uparrow}(j,q)$.  The eigenvalues are degenerate    and obey : $\epsilon(j=l;k)=\epsilon(j=-(l+1);k)$ ,where $l\geq0$. 
\begin{eqnarray}
&&q\equiv\frac{-i}{2 R_{g}(n)}\pm Q(\epsilon);\hspace{0.2 in}Q(\epsilon)=\sqrt{\epsilon^{2}-(\frac{l+\frac{1}{2}}{ R_{g}(n)})^2}\nonumber\\&&
F_{\epsilon}(l,q)\equiv[F_{\epsilon\uparrow}(l,q),F_{\epsilon\downarrow}(l,q)]^{T}\propto [1,e^{-i\kappa(Q,l)}]^{T};\hspace{0.1 in}  \kappa(Q,l)=tan^{-1}(\frac{Q R_{g}(n)}{l+\frac{1}{2}})\nonumber\\&& 
\end{eqnarray}
The value of the transversal momentum $ Q(\epsilon) $ will be determined  from the  boundary conditions at $\pm\frac{D(n)}{2}$. 
We will introduce a polar angle  $\theta$ measured with respect the Cartesian axes:
The  angle $0<\varphi(n=1)\leq 2\pi$  for  the upper contour  $n=1$  centered  at $[\overline{x}=0,\overline{y}=R_{g}]$ is described by the polar coordinate $ 0<\theta  \leq \pi$ measured from the center of the Cartesian coordinate $[0,0]$. The lower contour centered at $[\overline{x}=0,\overline{y}=-R_{g}]$  characterized by  the angle  $0<\varphi(n=-1)\leq 2\pi$  is described  by the polar angle $\theta$ restricted  to  $\pi<\theta \leq2\pi$. We establish the correspondence between  $\varphi(n=\pm1)$ and $\theta$:
\begin{eqnarray}
&&\varphi(n=1)=2\theta+\frac{3\pi}{2} \hspace{0.1 in}  for  \hspace{0.1 in}the \hspace{0.1 in} upper \hspace{0.1 in} contour \hspace{0.1 in} n=1, \hspace{0.1 in}  0<\theta  \leq \pi \nonumber\\&&
\varphi(n=-1)=2\theta+\frac{3\pi}{2}+\pi \hspace{0.1 in} for  \hspace{0.1 in} the \hspace{0.1 in} lower \hspace{0.1 in} contour \hspace{0.1 in} n=-1 , \hspace{0.1 in}0<\theta \leq\pi \nonumber\\&&
\end{eqnarray}
Following the discussion from the  previous chapter we will introduce the following boundary conditions:
\begin{eqnarray}  
&&F^{(n=1)}_{\epsilon\uparrow}(s,u=\frac{D}{2})=0;\hspace{0.4 in}
F^{(n=1)}_{\epsilon\downarrow}(s,y=-\frac{D}{2})=0 \nonumber\\&&
F^{(n=-1)}_{\epsilon\uparrow}(s,u=-\frac{D}{2})=0;\hspace{0.4 in}
F^{(n=-1)}_{\epsilon\downarrow}(s,y=\frac{D}{2})=0 \nonumber\\&&
D(n=\pm1)\equiv D
\end{eqnarray}
For the two contours $n=\pm1$ we  introduce eight spinors  
$\Gamma^{(n=\pm1)}_{l,Q}(\varphi(n=\pm1),u)$,$\Gamma^{(n=\pm1)}_{l,-Q}(\varphi(n=\pm1),u)$, $\Gamma^{(n=\pm1)}_{-l,Q}(\varphi(n=\pm1),u) $,  $\Gamma^{(n=\pm1)}_{-l,-Q}(\varphi(n=\pm1),u)$. Using this spinor  we will compute    the eigenfunctions: $F^{(n=1,R)}_{l,Q}(\varphi(n=1),u)$ ,   $ F^{(n=-1,L)}_{-l,Q}(\varphi(n=-1),u)$  characterized by the  transverse momentum $Q_{-}$ and the pair $F^{(n=-1,R)}_{l,Q}(\varphi(n=-1),u)$ , $F^{(n=1,L)}_{-l,Q}(\varphi(n=1),u)$ characterized by   the momentum $Q_{+}$.
\begin{equation*}
\Gamma^{(n=\pm1)}_{l,Q}(\varphi(n=\pm1),u)=e^{i l (\varphi(n=\pm1))}e^{iQ u} \left(\begin{array}{cc} 1\\ e^{i (\varphi(n=\pm1))} e^{-i\kappa(l,Q)}\end{array}
\right)
\end{equation*}
\begin{equation*}
\Gamma^{(n=\pm1)}_{l,-Q}(\varphi(n=\pm1),u)=e^{i l(\varphi(n=\pm1))}e^{-iQ u} \left(\begin{array}{cc} 1\\ e^{i (\varphi(n=\pm1))} e^{i\kappa(l,Q)}\end{array}
\right)
\end{equation*}
\begin{equation*}
\Gamma^{(n=\pm1)}_{-l,Q}(\varphi(n=\pm1),u)=e^{-i l(\varphi(n=\pm1))}e^{iQ u} \left(\begin{array}{cc} 1\\- e^{-i (\varphi(n=\pm1))} e^{i\kappa(l,Q)}\end{array}
\right)
\end{equation*}
\begin{equation}
\Gamma^{(n=\pm1)}_{-l,-Q}(\varphi(n=\pm1),u)=e^{-i l (\varphi(n=\pm1))}e^{-iQ u} \left(\begin{array}{cc} 1\\- e^{-i (\varphi(n=\pm1))} e^{-i\kappa(l,Q)}\end{array}
\right)
\label{spinors}
\end{equation}
Using the vanishing boundary condition given in equation $(40)$, we find for the  upper contour $n=1$ and positive angular momentum $l$ the wave functions:
\begin{eqnarray}
&&F^{(n=1,R)}_{\epsilon(l,Q_{-}),\uparrow}(\varphi(n=1),u)=e^{il \varphi(n=1)} e^{\frac{-i}{2}\kappa(l,Q_{-})}[e^{i(Q_{-}u+\frac{1}{2}\kappa(l,Q_{-}))}+(-1)^{k+1}e^{-i(Q_{-}u+\frac{1}{2}\kappa(l,Q_{-}))}] \nonumber\\&&
F^{(n=1,R)}_{\epsilon(l,Q_{-}),\downarrow}(\varphi(n=1),u)=e^{il \varphi(n=1)} e^{\frac{-i}{2}\kappa(l,Q_{-})} e^{i\varphi(n=1)}[e^{i(Q_{-}u-\frac{1}{2}\kappa(l,Q_{-}))}+(-1)^{k+1}e^{-i(Q_{-}u-\frac{1}{2}\kappa(l,Q_{-}))}]\nonumber\\&&
\end{eqnarray}
For the lower contour $n=-1$ and negative angular momentum  we find:
\begin{eqnarray}
&&F^{(n=-1,L)}_{\epsilon(-l,Q_{-}),\uparrow}(\varphi(n=-1),u)=e^{-il \varphi(n=-1)} e^{\frac{i}{2}\kappa(l,Q_{-})}[e^{i(Q_{-}u-\frac{1}{2}\kappa(l,Q_{-}))}+(-1)^{k+1}e^{-i(Q_{-}u-\frac{1}{2}\kappa(l,Q_{-}))}]\nonumber\\&&
F^{(n=-1,L)}_{\epsilon(-l,Q_{-}),\downarrow}(\varphi(n=-1),u)=e^{-il \varphi(n=-1)} e^{\frac{i}{2}\kappa(l,Q_{-})} (-1)e^{-i\varphi(-1)}[e^{i(Q_{-}u+\frac{1}{2}\kappa(l,Q_{-}))}\nonumber\\&&+(-1)^{k+1}e^{-i(Q_{-}u+\frac{1}{2}\kappa(l,Q_{-}))}]\nonumber\\&& 
\end{eqnarray}
For both set of eigenfunctions we obtain the  quantization conditions for the momentum  $Q_{-}$ and eigenvalues $\epsilon(l,Q_{-})$.
\begin{eqnarray}
&&Q_{-}=\frac{\pi}{D}k-\frac{1}{D}\tan^{-1}(\frac{Q_{-}R_{g}(1)}{l+\frac{1}{2}}),k=1,2,3...;
\tan[\kappa(l,Q_{-})]=(\frac{Q_{-}R_{g}(1)}{l+\frac{1}{2}})\nonumber\\&&
\epsilon(l,Q_{-})=\pm\hbar v_{F}\sqrt{(\frac{l+\frac{1}{2}}{ R_{g}(1)})^2+Q_{-}^2}
\end{eqnarray}
The presence of the singular transformation $e^{-\frac{ B^{(2)}}{2\pi}(\frac{1}{x+iy})}$ and $e^{-\frac{ B^{(2)}}{2\pi}(\frac{1}{x-iy})}$  demands that the eigenfunction should vanish  at $x=y=0$, which  corresponds to the point $u=0,\varphi(n=1)=-\frac{\pi}{2}$ for the upper contour and  $u=0,\varphi(n=-1)=\frac{\pi}{2}$ for the lower contour. Using the mapping  given eq.$(39)$ we can transform  from the variables $\varphi(n=\pm1)$ to the  polar coordinate  $\theta$ (the wave function must vanishes  at  $\theta=0$ and $\theta=\pi$).
In order to obtain a finite wave function, we combine the spinors  $F^{(n=1,R)}_{l,Q_{-}}(\varphi(n=1),u)$, $F^{(n=-1,L)}_{-l,Q_{-}}(\varphi(n=-1),u)$  with  the  singular transformation  functions $e^{-\frac{ B^{(2)}}{2\pi}(\frac{1}{x+iy})}$ , $e^{-\frac{ B^{(2)}}{2\pi}(\frac{1}{x-iy})}$:
\begin{eqnarray*}
U_{\epsilon(l,Q_{-})\uparrow}(\theta,u)&=&  e^{-\frac{ B^{(2)}}{2\pi}[(R_{g}+u)\cos[\varphi(n=1)]+i(R_{g}+ (R_{g}+u)\sin[\varphi(n=1)]]^{-1}}F^{(n=1,R)}_{l,Q_{-},\uparrow}(\varphi(n=1),u)\nonumber\\&&+ r(l,\uparrow) e^{-\frac{B^{(2)}}{2\pi}[(R_{g}+u)\cos[\varphi(n=-1)]+i(-R_{g}+ (R_{g}+u)\sin[\varphi(n=-1)]]^{-1}}F^{(n=-1,L)}_{-l,Q_{-},\uparrow}(\varphi(n=-1),u)
\end{eqnarray*}
and 
\begin{eqnarray*}
U_{\epsilon(l,Q_{-})\downarrow}(\theta,u)&=&e^{-\frac{ B^{(2)}}{2\pi}[(R_{g}+u)\cos[\varphi(n=1)]-i(R_{g}+ (R_{g}+u)\sin[\varphi(n=1)]]^{-1}}F^{(n=1,R)}_{l,Q_{-},\downarrow}(\varphi(n=1),u)\nonumber\\&&+ r(l,\downarrow)e^{-\frac{ B^{(2)}}{2\pi}[(R_{g}+u)\cos[\varphi(n=-1)]-i(-R_{g}+ (R_{g}+u)\sin[\varphi(n=-1)]]^{-1}}F^{(n=-1,L)}_{-l,Q_{-},\downarrow}(\varphi(n=-1),u)
\end{eqnarray*}
The amplitudes  $r(l,\uparrow)$ ,$r(l,\downarrow)$ are determined by  demanding the   vanishing  of the wave function  $U_{\epsilon(l,Q_{-})\uparrow}(\theta,u)$  and  $U_{\epsilon(l,Q_{-})\downarrow}(\theta,u)$ at $x=y=0$.  As a result, we obtain the explicit form of the wave functions  $U_{\epsilon(l,Q_{-})\uparrow}(\theta,u)$  $U_{\epsilon(l,Q_{-})\downarrow}(\theta,u)$  (see  equation $(76)$ in Appendix-C)  which depend  of the parameters $\eta(u)$ and $\zeta(\theta,u)$ :   $\eta(u)=\frac{R_{g}(1)}{R_{g}(1)+u},\frac{|u|}{R_{g}(1)}<1;\hspace{0.2 in}\zeta(\theta,u)=\frac{-B^{(2)}}{2\pi(R_{g}(1)+u)((\sin[2\theta])^2+(\eta(u)-\cos[2\theta])^2)}$.

For the second pair  $F^{(n=-1,R)}_{l,Q_{+}}(\varphi(n=-1),u)$ , $F^{(n=1,L)}_{-l,Q_{+}}(\varphi(n=1),u)$ we obtain  the quantization conditions: $Q_{+}=\frac{\pi}{D}k+\frac{1}{D}\tan^{-1}(\frac{Q_{+}R_{g}(1)}{l+\frac{1}{2}})$, k=1,2,3... $\tan[\kappa(l,Q_{+})]=(\frac{Q_{+}R_{g}(1)}{l+\frac{1}{2}})$
with the eigenvalues: $\epsilon(l,Q_{+})=\pm\hbar v_{F}\sqrt{(\frac{l+\frac{1}{2}}{ R_{g}(n)})^2+Q_{+}^2}$.
Following the same procedure as we used for  the  pairs  $F^{(n=1,R)}_{l,Q_{-}}(\varphi(n=1),u)$ ,   $ F^{(n=-1,L)}_{-l,Q_{-}}(\varphi(n=-1),u)$ 
we obtain  the wave  functions   $U_{\epsilon(l,Q_{+})\uparrow}(\theta,u)$ , $U_{\epsilon(l,Q_{+})\downarrow}(\theta,u)$  which are  given in  equation $(77)$ Appendix-C.

\vspace{0.2 in}

\textbf{VI -Computation of the  STM  density of states}

\vspace{0.2 in}

The STM tunneling current $I$ is a function of the bias voltage $V$ which  gives spatial and spectroscopic information about the  electronic surface states. At zero temperature, the derivative of the current with respect the bias voltage  $V$ is given in term of the single particles eigenvalues: $\epsilon(m,q_{-})=\pm\hbar v_{F}\sqrt{(\frac{2\pi}{L}m)^2+q_{-}^2}$,   $\epsilon(m,q_{+})=\pm\hbar v_{F}\sqrt{(\frac{2\pi}{L}m)^2+q_{-}^2}$ ,$m=0,1,2,3...$  for    contour $n=0$.
For the upper and lower circular contours $n=\pm1$,  we have :$\epsilon(l,Q_{-})=\pm\hbar v_{F}\sqrt{(\frac{l+\frac{1}{2}}{ R_{g}(1)})^2+Q_{-}^2}$ ,$\epsilon(l,Q_{+})=\pm\hbar v_{F}\sqrt{(\frac{l+\frac{1}{2}}{ R_{g}(1)})^2+Q_{+}^2}$  ,$l=0,1,2,3..$.
The $STM$ density of states  is computed for a voltage $V$ between the $STM$ tip and the sample. The  tunneling current is   a function of the   bias voltage  $V$  and the  chemical potential  $\mu>0$  \cite{kittel}: 
\begin{eqnarray}
&&\frac{d I}{dV}\propto D(E=eV;s,u)\equiv \sum_{n} D^{(n)}(E=eV;s,u)= \nonumber\\&&= \sum_{\eta=\pm}[\sum_{m}\sum_{q_{r}=q_{+},q_{-}}\sum_{\sigma}|U^{(n=0;m,q_{r})}_{\sigma}(x,y)|^{2}\delta[eV +\mu -\eta\hbar v_{F}\sqrt{(\frac{2\pi}{L}m)^2+q_{r}^2}]\nonumber\\&&+\sum_{n=\pm1}\sum_{l}\sum_{Q_{r}=Q_{+},Q_{-}}\sum_{\sigma}|U^{(n=\pm1;l,Q_{r})}_{\sigma}(\theta,u)|^{2}\delta[eV +\mu -\eta\hbar v_{F}\sqrt{(\frac{l+\frac{1}{2}}{ R_{g}(1)})^2+Q_{r}^2}]]\nonumber\\&&
\end{eqnarray}
($\eta=+$ corresponds to electrons  with energy $0<\epsilon\leq \mu$ and  $\eta=-$ corresponds to electrons below the Dirac point $\epsilon<0$.  For the  rest   of this   paper  we will take the chemical potentials to be  $\mu= 120mV$  (this is typical value for the $TI$ ). We will neglect   the states with $\eta=-$ which correspond to particles below the Dirac cone ( see Appendix -B).
The   density of states at the  tunneling energy $eV$  is weighted  by the probability density of  the $STM$ tip  at position $[x,y]$  for n=0. The contours for  $n=\pm1$    will be parametrized  in terms of  the polar angle  $\theta$ and transverse coordinate $u$.
The proportionality factor $J$ for  the tunneling probability (not shown  in the equation )  $\frac{d I}{dV}=J  D(V;x,y)$  is a function of the distance between the tip and the sample.  The notation   $D^{(n)}(V;x,y)$ represents  the tunneling  density for the different contours.

\vspace{0.2 in}

\textbf{A- The  tunneling density of states  for the    Peierls  model $D^{Peierls}(V;x,y)$  }

\vspace{0.2 in}

We consider first  the  Peierls model introduced in section $III-A$. This model has a zero mode  for which we have computed  the  wave function  in equation $(6)$.
We find that the tunneling density of states density is   confined   to  the quantum strip   $y=0$ with a varying    width  $\widetilde{W}$, $|y|< \widetilde{W}$ determined by the dislocations distribution \cite{kosevich} 
\begin{eqnarray}
&&D^{Peierls}(V;\frac{x}{\widetilde{W}},\frac{y}{\widetilde{W}})\propto (\frac{L}{h v_{F}})^2(\frac{B^{(2)}}{L})^2 e^{-\frac{2\kappa B^{(2)}}{\pi}  \int_{0}^{|y|}\,dy' \tan^{-1}(\frac{y'}{\widetilde{W}})}\nonumber\\&&
\end{eqnarray} 
In figure $3$ we show the tunneling density of states for the parameters $\frac{2\kappa B^{(2)}}{\pi}=0.1$ in units of the half width  $\widetilde{W}$.

\vspace{0.2 in}

\textbf{B-The  tunneling density of states $D^{(n=0)}(V;x,y)$  for   $n=0$}

\vspace{0.2 in}

Summing up the single particle states  weighted with occupation  probability  $|U^{(n=0;m,q_{r})}_{\sigma}(x,y)|^{2}$, we  obtain a space dependent  density of states for the  two dimensional boundary surface ,$\frac{-L}{2}\leq x\leq\frac{L}{2}$ and the coordinate $y$ is restricted to the regions $\frac{d}{2}<y\leq \frac{L}{2}$  and $\frac{-L}{2}<y\leq \frac{d}{2}$.  
We will perform the computation at the thermodynamic limit, namely we replace the discrete momentum $\frac{\pi}{L}k$ by $Y=\frac{k}{N}$  and $\frac{2\pi}{L}m$ by $X= \frac{m}{N}$ where $N=\frac{L}{a}$. We find for  the dimensionless momentum $\hat{q}\equiv q a$  the equations :
$ \hat{q}_{\pm}(Y)=\pi Y \pm\frac{1}{N}\tan^{-1}[\frac{\hat{q}_{\pm}(Y)}{2\pi X}]$ where   $2\pi X=pa=\hat{p}$.  As a result we obtain the following density of states $\frac{\partial \hat{q}_{\pm}}{\partial Y}$
\begin{eqnarray} 
&&[\frac{\partial \hat{q}_{+}}{\partial Y}]^{-1}=\frac{1}{\pi}\frac{\hat{q}_{+}^{2}+\hat{p}^{2}-\frac{1}{N}\hat{p}}{\hat{q}_{+}^{2}+\hat{p}^{2}}\nonumber\\&&
[\frac{\partial \hat{q}_{-}}{\partial Y}]^{-1}=\frac{1}{\pi}\frac{\hat{q}_{-}^{2}+\hat{p}^{2}+\frac{1}{N}\hat{p}}{\hat{q}_{-}^{2}+\hat{p}^{2}}\nonumber\\&&
\end{eqnarray}
Using this results, we  compute  the tunneling density of states in terms of the  energy  $\mu+eV$  measured with respect the chemical potential $\mu$ and  the  transverse energy $\epsilon_{\bot}\equiv \hbar v_{F}q_{\pm}$.
\begin{eqnarray}
&&D^{(n=0)}(V;x,y)= (\frac{L}{h v_{F}})^{2}(\frac{B^{(2)}}{L})^2   \frac{1}{4\sqrt{G(x,y)}}e^{\frac{-B^{(2)}}{\pi}(\frac{x}{x^2+y^2+a^2})}[\int_{0}^{E_{max.}}\,d \epsilon_{\bot}\frac{(\mu+eV)}{\sqrt{(\mu+eV)^2-\epsilon_{\bot}^2}}\dot\nonumber\\&&[\frac{1}{2}(1+\frac{1}{\pi}\frac{hv_{F}}{L (\mu+V)}\sqrt{1-(\frac{\epsilon_{\bot} }{\mu+V})^2}\hspace{0.1 in})+ \frac{1}{2}(1-\frac{1}{\pi}\frac{hv_{F}}{L (\mu+V)}\sqrt{1-(\frac{\epsilon_{\bot} }{\mu+V})^2}\hspace{0.1 in})] \nonumber\\&&+
\frac{h v_{F}}{L}(H[\mu+V-\frac{h v_{F}}{2 L}]-H[\mu+eV-E_{max}])\cdot((\cos[\frac{(\mu+eV)}{\hbar v_{F}}y-\frac{\pi}{4}])^2+(\cos[\frac{(\mu+eV)}{\hbar v_{F}}y-\frac{\pi}{4}])^{2})]\nonumber\\&&=
(\frac{L}{h v_{F}})^{2}(\frac{B^{(2)}}{L})^2   \frac{1}{4\sqrt{G(x,y)}}e^{\frac{-B^{(2)}}{\pi}(\frac{x}{x^2+y^2+a^2})}[\int_{0}^{E_{max.}}\,d \epsilon_{\bot}\frac{(\mu+eV)}{\sqrt{(\mu+eV)^2-\epsilon_{\bot}^2}} \nonumber\\&&+
\frac{h v_{F}}{L}(H[\mu+V-\frac{h v_{F}}{2 L}]-H[\mu+eV-E_{max}])\cdot((\cos[\frac{(\mu+eV)}{\hbar v_{F}}y-\frac{\pi}{4}])^2+(\cos[\frac{(\mu+eV)}{\hbar v_{F}}y-\frac{\pi}{4}])^{2})]=\nonumber\\&&
(\frac{L}{h v_{F}})^{2}(\frac{B^{(2)}}{L})^2   \frac{1}{4\sqrt{G(x,y)}}e^{\frac{-B^{(2)}}{\pi}(\frac{x}{x^2+y^2+a^2})}[\frac{\pi}{2}(\mu+eV)+\frac{h v_{F}}{L}(H[\mu+V-\frac{h v_{F}}{2 L}]-H[\mu+eV-E_{max}])]\nonumber\\&&for\hspace{0.2 in}  |y|>d
\nonumber\\&&
\end{eqnarray}
$H[\mu+eV-\frac{h v_{F}}{2 L}]$ is the step function  which is one for  $\mu+eV-\frac{h v_{F}}{2 L}\geq 0$ and  zero otherwise. $a=\frac{2\pi}{\Lambda}$ is the short distance cut-off and $E_{max}=\hbar v_{F}\Lambda<0.3 eV$ is the maximal energy  which restricts the validity of the Dirac model.  We observe  in the second  line that the  asymmetry in the density of states  $1\pm\frac{1}{\pi}\frac{hv_{F}}{L (\mu+V)}\sqrt{1-(\frac{\epsilon_{\bot} }{\mu+V})^2}\hspace{0.1 in})$ cancels.  

Equation $(48)$   shows  that the tunneling density of states is linear in the energy   $\mu+eV$  (in the present case we have looked only for energies above the Dirac cone ). For the   chemical potential $\mu=120mV$,  the zero energy corresponds to the Voltage $V=-120 mV$. The tunneling density of states has a constant part at energies $\frac{h v_{F}}{2 L}\approx 0.2 mV$  for $-120 mV<V<-119.8mV$. For $V>-119.8mV$ the density of states is proportional to $\mu+eV$.

In figure $4$ we have plotted the tunneling density of states as a function of the coordinates $x$ and $y$. The shape of the plot is governed by the the multiplicative  factor $e^{-\frac{B^{(2)}}{\pi}(\frac{x}{x\pm i y})}$ which governs the solutions in eq.$(35)$. We observe that the density of state is  maximal in the region $|y|<10 B^{(2)} $. 

Figure $5$ shows the dependence on the voltage $V$ and coordinate $y$. We observe the linear increase in the tunneling density of states which is maximal  in the region $|y|<10 B^{(2)}$.

\vspace{0.2 in} 

\textbf{C-The tunneling  density of states $D^{(n=0)}(V,x,y;\vec{r}_{1},..\vec{r}_{2M})$  for $2M$ dislocations.}

\vspace{0.2 in}

For many dislocations which  satisfy  $\sum_{w=1}^{2M}B^{(2,w)}=0$  ( sum of the Burger vectors is zero ) with  the core  centered at $[x_{w},y_{w}]$ ,$w=1,2..2M$ the coordinate $\vec{r}=(x,y)\rightarrow [X(\vec{r}),Y(\vec{r})]$ is replaced by  $[X(\vec{r})=x,Y(\vec{r})=y+\sum_{w}\frac{B^{(2,w)}}{2\pi}\tan^{-1}(\frac{y-y_{w}}{x-x_{w}})]$.
Following the method used previously,  we find the edge Hamiltonian with many dislocations  takes the form: 
\begin{equation}
h^{edge}(w=1,2...2M)\approx i\sigma^{1}[\partial_{y} -\frac{i}{2} \sum_{w=1}^{2M}\sigma^{3}B^{(2,w)}\delta^{2}(\vec{r}-\vec{r}_{w})] -i\sigma^{2}\partial_{x}
\label{Burgermany}
\end{equation}
As a result, the wave functions are given by: 
\begin{eqnarray}
&&U^{(n=0,w=1,2...2M)}_{\uparrow}(x,y)\propto
\prod_{w=1,2...2M}e^{-\frac{ B^{(2)}}{2\pi}(\frac{1}{(x-x_{w})+i(y-y_{w})})}F^{(n=0)}_{\uparrow}(x,y)\nonumber\\&&
U^{(n=0,w=1,2...2M)}_{\downarrow}(x,y)\propto
\prod_{w=1,2...2M}e^{-\frac{ B^{(2)}}{2\pi}(\frac{1}{(x-x_{w})-i(y-y_{w})})}F^{(n=0)}_{\downarrow}(x,y)
\end{eqnarray}
Using these wave functions, we find that the tunneling density of states is given by:
\begin{equation}
D^{(n=0)}(V,x,y;\vec{r}_{1},..\vec{r}_{2M})\propto \prod_{w=1,2...2M}e^{-\frac{ B^{(2)}}{\pi}(\frac{(x-x_{w})}{(x-x_{w})^2+(y-y_{w})^2+a^2})}
\label{densityw}
\end{equation}
In figure $6$ we show the tunneling density of states for an even number of dislocations in the $y$ directions which have the core on the $y=0$ axes ($\vec{r}_{w}=[x_{w},y_{w}=0]$, $w=1,2,3,...2M$).
We observe that the tunneling density of states is confined to the axes $y=0$ and resembles the structure obtained from the  Peierls  model given in figure 3.

\vspace{0.2 in} 

\textbf{D-The tunneling  density of states $D^{(n=\pm1)}(V;\theta,u)$  for the $n=\pm1 $ contours.}

\vspace{0.2 in}

Following the  same procedure  as used for the $n=0$ case  and the eigenfunction  given in Appendix-C, we find for the tunneling density of states:
\begin{equation}
D^{(n=\pm1)}(V;\theta,u)\equiv D^{(n=\pm1)}(\mu,V;\theta,u)_{even}+ D^{(n=\pm1)}(\mu,V;\theta,u,\mu)_{odd}
\label{contoursd}
\end{equation}
For the even $k$'s, we solve for the momentum $Q_{+}$ and  $Q_{-}$ and find:
\begin{eqnarray}
&&D^{(n=\pm1)}(\mu,V;\theta,u)_{even}=\frac{(B^{(2)})^{2}}{2\pi R_{g}(1) D(1) \sqrt{G(\theta,u)}}\sum_{Q_{r}=Q_{+},Q_{-}}\sum_{l=0}^{\infty}\delta[eV+\mu-\hbar v_{F}\sqrt{(\frac{l+\frac{1}{2}}{ R_{g}(1)})^2+Q_{r}^2}]\nonumber\\&&
[(e^{-2\zeta(\theta,u)\sin[2\theta]}+e^{2\zeta(\theta,u)\sin[2\theta]})((\sin[Q_{r}u-\frac{1}{2}\kappa(l,Q_{r})])^2+(\sin[Q_{r}u+\frac{1}{2}\kappa(l,Q_{r})])^2)+\nonumber\\&& 2(-1)^{l}\sin[Q_{r}u+\frac{1}{2}\kappa(l,Q_{r})]\sin[Q_{r}u-\frac{1}{2}\kappa(l,Q_{r})]\cdot\nonumber\\&&(
\cos[l(\theta+\frac{3\pi}{2}) -\zeta(\theta,u)(-\eta(u)+\cos[2\theta])]-\cos[(l+1)(\theta+\frac{3\pi}{2}) +\zeta(\theta,u)(-\eta(u)+\cos[2\theta])]];\nonumber\\&&
\end{eqnarray}
Similarly for the odd $k$'s we find:
\begin{eqnarray}
&&D^{(n=\pm1)}(\mu,V;\theta,u)_{odd}=\frac{(B^{(2)})^{2}}{2\pi R_{g}(1) D(1) \sqrt{G(\theta,u)}} \sum_{Q_{r}=Q_{+},Q_{-}}\sum_{l=0}^{\infty}\delta[eV+\mu-\hbar v_{F}\sqrt{(\frac{l+\frac{1}{2}}{ R_{g}(1)})^2+Q_{r}^2}]\nonumber\\&&
[(e^{-2\zeta(\theta,u)\sin[2\theta]}+e^{2\zeta(\theta,u)\sin[2\theta]})((\cos[Q_{r}u-\frac{1}{2}\kappa(l,Q_{r})])^2+(\cos[Q_{r}u+\frac{1}{2}\kappa(l,Q_{r})])^2)+\nonumber\\&& 2(-1)^{l}\cos[Q_{r}u+\frac{1}{2}\kappa(l,Q_{r})]\cos[Q_{r}u-\frac{1}{2}\kappa(l,Q_{r})]\cdot\nonumber\\&&(
\cos[l(\theta+\frac{3\pi}{2}) -\zeta(\theta,u)(-\eta(u)+\cos[2\theta])]-\cos[(l+1)(\theta+\frac{3\pi}{2}) +\zeta(\theta,u)(-\eta(u)+\cos[2\theta])]]\nonumber\\&&
\end{eqnarray}
For the present case the energy scale of the excitations is governed by the radius $R_{g}(1)$ and width $D$. The spectrum is discrete and we can't replace it by a continuum density of states as we did for the case $n=0$.

In figure $7$ we show the tunneling density of states at a fixed polar angle $\theta=\frac{\pi}{2}$  as a function of the voltage $V$. We observe that the density of states is dominated by high energy eigenvalues. This solutions are localized in energy. The range of the spectrum is above  $\mu+eV>200mV$ which  is well separated from  the low energy spectrum controlled by the $n=0$ contour (which ranges from $-120 mV $ to $70 mV$).

 Figure $8$   shows the  tunneling density of states as a function of the polar angle $\theta$ for a fixed energy . The periodicity in $\theta $  is controlled  by the discrete energy   eigenvalues. 

In figure  $9$  we show  the tunneling density of states  at a fixed voltage $V$  as a function of the polar angle $0<\theta<\pi$ and width $|u|<0.1 $.

\vspace{0.2 in}

\textbf{VII-The  charge current-the in plane spin on the  surface}

\vspace{0.2 in}

\textbf{A-The   current in the absence of the edge dislocation}

\vspace{0.2 in}

From the Hamiltonian given in equation $1$ we  compute the equation of motion for the velocity operator:
 $\frac{dx}{dt}=\frac{1}{i\hbar}[x,h]=v_{F}\sigma^{y}$ , $\frac{dy}{dt}=\frac{1}{i\hbar}[y,h]=-v_{F}\sigma^{x}$.   We  multiply the velocity operator  by the charge $(-e)$ and  identify the charge current operators :
$\hat{J}_{x}=(-e)v_{F}\sigma^{2}$,  $\hat{J}_{y}=(-e)(-v_{F})\sigma^{1}$.
This  also represent the "`real"' spin on the surface. Therefore, the charge current is a measure of the in-plane spin on the  surface.

Integrating over the $y$ coordinate we obtain the current $I_{x}^{T.I.}$ in the $x$ direction. Using the  eigenstates   $\Gamma_{p,q}(x,y)$  and $\Gamma_{-p,q}(x,y)$ of the $h^{T.I.}$ Hamiltonian 
\begin{equation*}
\Gamma_{p,q}(x,y)=e^{i p x}e^{iq y} \left(\begin{array}{cc} 1\\ i e^{i\chi(p,q)}\end{array}
\right)
\end{equation*}
\begin{equation*}
\Gamma_{-p,q}(x,y)=e^{-i p x}e^{iq y} \left(\begin{array}{cc} 1\\ -i e^{-i\chi(p,q)}\end{array}
\right)
\end{equation*}
we find $(\Gamma_{p,q}(x,y))(\sigma^{2})(\Gamma_{p,q}(x,y))=- (\Gamma_{-p,q}(x,y))(\sigma^{2})(\Gamma_{-p,q}(x,y))$ therefore, we conclude that the current $I_{x}^{T.I.}=0$ is zero.

\vspace{0.2 in}

\textbf{B-The  current  in the presence of the edge dislocation}

\vspace{0.2 in}

We  will compute the current in the presence  of the edge dislocation.
The current operator $\hat{J}^{edge}_{x}(x,y)$  will be given in terms of the transformed currents. We find that the  current  density operator $J^{edge}_{x}(x,y)$ is given by:
\begin{equation}
\hat{J}^{edge}_{x}(x,y)=(-e)v_{F}[\sigma^{2}e^{x}_{1}-\sigma^{1}e^{x}_{2}]= (-e)v_{F}\sigma^{2}-(-e)v_{F}\frac{B^{(2)}}{2\pi}(\frac{ y\sigma^{1}+ x\sigma^{2}}{x^2+y^2})\approx (-e)v_{F}\sigma^{2}
\label{edge}
\end{equation}
We use the zero order  current operator  $\hat{J}^{edge}_{x}(x,y)\approx (-e)v_{F}\sigma^{2}$ to construct  the second quantization form for  the current  density. The operator  is defined with respect the to  shifted ground state $|\mu>\equiv|\tilde{0}>$  with the energy $E=\epsilon-\mu$ measured with respect the chemical potential and spinor  field $\Psi_{n=0}(x,y)$.
\begin{equation}
J^{edge}_{x}(x,y)= <\mu|\Psi_{n=0}^{\dagger}(x,y)\hat{J}^{edge}_{x}(x,y)\Psi_{n=0}(x,y)|\mu>
\label{spinor}
\end{equation} 
 Using the spinor eigenfunction given in equation $(35)$ and  the second quantized form given in Appendix -B  we find :
\begin{equation*}
\Psi_{n=0}(x,y;t)\approx\sum_{E>0} [\alpha_{E,R}\left(\begin{array}{cc}U^{(n=0,R)}_{\uparrow}(x,y)\\U^{(n=0,R)}_{\downarrow}(x,y)\end{array}
\right)_{E+\mu}{e^{-i\frac{E}{\hbar}t}}+\beta^{\dagger}_{E,R} \left(\begin{array}{cc}U^{(n=0,R)}_{\uparrow}(x,y)\\U^{(n=0,R)}_{\downarrow}(x,y)\end{array}
\right)_{-E+\mu}e^{i\frac{E}{\hbar}t}
\end{equation*}
\begin{equation} +\alpha_{E,L}\left(\begin{array}{cc}U^{(n=0,L)}_{\uparrow}(x,y)\\U^{(n=0,L)}_{\downarrow}(x,y)\end{array}
\right)_{E+\mu}e^{-i\frac{E}{\hbar}t}+\beta^{\dagger}_{E,L} \left(\begin{array}{cc}U^{(n=0,L)}_{\uparrow}(x,y)\\U^{(n=0,L)}_{\downarrow}(x,y)\end{array}
\right)_{-E+\mu}e^{i\frac{E}{\hbar}t}]
\label{eqsp}
\end{equation} 
The current is a sum  of two terms computed with the eigen spinor obtained in equation $(33)$: $[U^{(n=0,R)}_{\uparrow}(x,y),U^{(n=0,R)}_{\downarrow}(x,y)]^{T}  \sigma^{2}[U^{(n=0,R)}_{\uparrow}(x,y),U^{(n=0,R)}_{\downarrow}(x,y)]$ 
 and  
 $[U^{(n=0,L)}_{\uparrow}(x,y),U^{(n=0,L)}_{\downarrow}(x,y)]^{T}  \sigma^{2}[U^{(n=0,L)}_{\uparrow}(x,y),U^{(n=0,L)}_{\downarrow}(x,y)]$ 
which have  opposite signs.  Due to the parity violation caused by the dislocation,   the density of states is asymmetric    $1\pm\frac{1}{\pi}\frac{hv_{F}}{L (\mu+V)}\sqrt{1-(\frac{\epsilon_{\bot} }{\mu+V})^2}\hspace{0.1 in})$ resulting in a finite current.  We 
integrate over the transversal direction  $y$ and obtain the  edge current  $I_{x}^{n=0,edge}$.
\begin{eqnarray}
&&I^{n=0,edge}= (-e)v_{F}\int_{-\frac{L}{2}}^ {\frac{L}{2}}\frac{d x}{L}\int_{-\frac{L}{2}}^{\frac{L}{2}}\,dy<\mu|J^{edge}(x,y)|\mu>=\nonumber\\&&
\frac{(-e)v_{F}}{4\pi}(\frac{L}{h v_{F}})^2(\frac{1}{L})\int_{-\frac{L}{2}}^ {\frac{L}{2}}\frac{d x}{L}\int_{-\frac{L}{2}}^{\frac{L}{2}}\frac{d y}{L}\frac{ e^{\frac{-B^{(2)}}{\pi}(\frac{x}{x^2+y^2+a^2})}}{\sqrt{G(x,y)}}
\int\,d\epsilon_{||}\int d\epsilon_{\bot}H[\mu-\sqrt{(\epsilon_{||})^2+(\epsilon_{\bot})^2}\hspace{0.1 in}]\frac{(hv_{F}/L)\cdot \epsilon_{||}}{(\epsilon_{||})^2+(\epsilon_{\bot})^2}\nonumber\\&&
=\frac{1}{4\pi}(\frac{-e v_{F}}{L})(\frac{\mu}{hv_{F}/L})f[\frac{B^{(2)}}{L}]\cdot(H[\mu+eV-\frac{hv_{F}}{L}]-H[\mu+eV-E_{max.}]);\hspace{0.2 in}
f[\frac{B^{(2)}}{L}]\approx 6.22\nonumber\\&&
\end{eqnarray}
$H[\mu-\sqrt{(\epsilon_{||})^2+(\epsilon_{\bot})^2}\hspace{0.1 in}]$ is the step function which is one for $\sqrt{(\epsilon_{||})^2+(\epsilon_{\bot})^2}\leq \mu$. The single particle energies are $\epsilon_{\bot}=\hbar v_{F}q_{\pm}$  and $\epsilon_{||}=\hbar v_{F}p$.
For $L\approx 10^{-6}m $, chemical potential  $\mu=120mV$   and $\frac{L}{B^{(2)}}\approx 100$ we find  that the current  $I_{x}^{n=0,edge}$ is in the range of $mA$.

To conclude, we have shown that the presence of an edge dislocation gives rise to a non-zero current which is a manifestation of the in-plane component of the spin on the two dimensional  surface . Therefore a  nonzero value  $I_{x}^{n=0,edge}\neq 0$ will be an indication of the presence of the edge dislocation. This  effect might be measured using  a coated tip with magnetic material used by the technique of Magnetic Force Microscopy.

\vspace{0.2 in}

\textbf{VIII-Conclusions}

\vspace{0.2 in}

Using the full description of the edge dislocation in terms of the torsion  tensor, we have shown that the singularity at the core center eliminates the zero mode . As a results only weak backscattering effect is allowed . 
Using this formulation we have shown that  for the case $n=0$ the tunneling density of states is confined along $y=0$. At  high energies the tunneling density of states  is confined to circular contours governed by the  Burger vector .  For a large  number of dislocations  we obtain a result similar to the one obtained from the Peierls domain wall model. 
The in plane spin orientation is a manifestation of the  parity violation  induced by the  edge dislocation.
We propose that scanning tunneling  and Magnetic Force Microscopy  are advanced  experimental techniques which   can  verify our predictions.

\pagebreak
  
\vspace{0.2 in}

\textbf{ Appendix -A}

\vspace{0.2 in}

We consider that    a two dimensional   manifold with  a   mapping  from the curved space   $X^{a}$, $ a=1,2 $,   to the $local $ $flat$ space   $x^{\mu}$,     $\mu=x,y$ exists.
We introduce the tangent vector \cite{Green}
 $e^{a}_{\mu}(\vec{x})=\frac{\partial X^{a}(\vec{x})}{\partial x^{\mu}}$,  $\mu=x,y$
 which satisfies the orthonormality relation  $e^{a}_{\mu}(\vec{x})e^{b}_{\mu}(\vec{x})=\delta_{a,b}$  (here we use the convention that we sum over indices which appear twice).  The metric tensor  for the curved space is given in terms of the flat metric $\delta_{a,b}$ and the scalar product of the tangent vectors: $e^{a}_{\mu}(\vec{x})e^{a}_{\nu}(\vec{x})=g_{\mu,\nu}(\vec{x})$.
The linear connection is determined by the Christoffel   tensor  $ \Gamma^{\lambda}_{\mu,\nu} $ :
  
\begin{equation}
 \nabla_{\partial_{\mu}}\partial_{\nu}= -\Gamma^{\lambda}_{\mu,\nu} \partial_{\lambda}
 \label{christ}
 \end{equation}
 
The Christoffel tensor is constructed from the metric tensor $ g_{\mu,\nu}(\vec{x})$.
\begin{equation}
\Gamma^{\lambda}_{\mu,\nu}=-\frac{1}{2}\sum_{\tau=x,y}g^{\lambda,\tau}(\vec{x})[\partial_{\nu}g_{\nu,\tau}(\vec{x})+\partial_{\mu}g_{\nu,\tau}(\vec{x})-\partial_{\tau}g_{\mu,\nu}(\vec{x})]
\label{gama}
\end{equation}

Next, we introduce the vector  field $\vec{V}=V^{a}\partial_{a}=V^{\mu}\partial_{\mu}$  where $a=1,2$  are the components in the curved space and $\mu=x,y$ represents the coordinate in the fixed cartesian  frame. The covariant derivative of the vector field  $V^{a}$ is determined by the  spin connection $\omega_{q,b}^{\mu}$ which needs to be computed: 

\begin{equation}
D_{\mu}V^{a}(\vec{x})=\partial_{\mu}V^{a}(\vec{x})+\omega_{a,b}^{\mu} V^{b}
\label{vector}
\end{equation}

For a two component spinor, we can identify the spin connection in the following way:   The spinor in the  the curved space (generated by the dislocation) is represented by $\widetilde{\Psi}(\vec{X})$  and  in the Cartesian space it is given by  is given  by  $\Psi(\vec{x})$   \cite{Maggiore}.
 The two component spinor represents a chiral fermion  which transform under spatial rotation as  spin half fermion:  

\begin{eqnarray}
&&\widetilde{\Psi}(\vec{X})=e^{\frac{-i}{2}\omega_{1,2}\sigma_{3}}\Psi(\vec{x})\nonumber\\&& e^{\frac{-i}{2}\omega_{1,2}\sigma_{3}}\equiv e^{\frac{1}{2}\omega_{a,b}\Sigma^{a,b}}\equiv e^{\sum_{a=1,2}\sum_{b=1,2}\frac{1}{2}\omega_{a,b}\Sigma^{a,b}}\nonumber\\&&
\omega_{a,b}\equiv-\omega_{b,a}\nonumber\\&&
\Sigma^{a,b}\equiv \frac{1}{4}[\sigma^{a},\sigma^{b}]\nonumber\\&&
\end{eqnarray}
 We have used the anti symmetric property of the rotation matrix  $\omega_{a,b}\equiv-\omega_{b,a}$, and the representation of the  generator $\Sigma^{a,b}$ in terms of the Pauli matrices. 
 
Therefore for a two component spinor we obtain the connection:

\begin{equation}
D_{\mu}\Psi(\vec{x})=(\partial_{\mu}+\frac{1}{2}\omega^{a,b}_{\mu}\Sigma_{a,b})\Psi(\vec{x})\equiv (\partial_{\mu}+\frac{1}{8}\omega^{a,b}_{\mu}[\sigma_{a},\sigma_{b}])\Psi(\vec{x})
\label{tensor}
\end{equation}
%With the anti-symmetric property $\omega^{a,b}_{\mu}=-\omega^{b,a}_{\mu}$.

Next we will compute the spin connection  $\omega^{a,b}_{\mu}$ using the \textbf{Christoffel tensor}.
 In the physical coordinate  basis $x^{\mu}$  the covariant derivative  $D_{\mu}V^{\nu}(\vec{x})$ is determined by the Christoffel tensor: 

\begin{equation}
D_{\mu}V^{\nu}(\vec{x})=\partial_{\mu}V^{\nu}(\vec{x})+\Gamma^{\lambda}_{\mu,\nu}  V^{\lambda}
\label{connection}
\end{equation}

The relation between the spin connection and the linear connection can be obtained from the fact that the two covariant derivative of the vector $\vec{V}$ are equivalent.

\begin{equation}
D_{\mu}V^{a}=e^{a}_{\nu}D_{\mu}V^{\nu}
\label{rel}
\end{equation}

Since we have the relation $V^{a}=e^{a}_{\nu}V^{\nu}$  it follows from the last equation 

\begin{equation}
D_{\mu}[e^{a}_{\nu}]=D_{\mu}\partial_{\nu}e^{a}=(D_{\mu}\partial_{\nu})e^{a}+ \partial_{\nu}(D_{\mu}e^{a})=0
\label{relt}
\end{equation}

Using the definition of the Christoffel index  and the differential geometry relation $\nabla_{\partial_{\mu}}\partial_{\nu}= -\Gamma^{\lambda}_{\mu,\nu} \partial_{\lambda}$
\cite{Green},  we obtain  the relation between the spin connection and the linear connection:

\begin{equation}
D_{\mu}[e^{a}_{\nu}]=\partial_{\mu}e^{a}_{\nu}(\vec{x})-\Gamma^{\lambda}_{\mu,\nu} e^{a}_{\lambda}(\vec{x})+ \omega^{a}_{\mu,b}e^{b}_{\nu}(\vec{x})\equiv0
\label{dd}
\end{equation}

Solving this  equation, we obtain  the spin connection  given in terms of the Burger vector. 
We multiply from left equation  by the tangent  vector $e^{a}_{\nu}$, replace  $\Gamma^{\lambda}_{\mu,\nu}$ by equation $(54)$ use the  metric tensor relations  $e^{a}_{\mu}(\vec{x})e^{b}_{\mu}(\vec{x})=\delta_{a,b}$,  $e^{a}_{\mu}(\vec{x})e^{a}_{\nu}(\vec{x})=g_{\mu,\nu}(\vec{x})$. 
As a result, we find \cite{Green}:
\begin{eqnarray} &&\omega^{a,b}_{\mu}=\frac{1}{2}e^{\nu,a}(\partial_{\mu}e^{b}_{\nu}-\partial_{\nu}e^{b}_{\mu})-
\frac{1}{2}e^{\nu,b}(\partial_{\mu}e^{a}_{\nu}-\partial_{\nu}e^{a}_{\mu})\nonumber\\&&
-\frac{1}{2}e^{\rho,a}e^{\sigma,b}(\partial_{\rho}e_{\sigma, c}-\partial_{\sigma}e_{\rho, c})e^{c}_{\mu}
\end{eqnarray}
We notice the asymmetry   between $e^{\nu,a}$ and  $e_{a,\nu}$:
 
$e^{\nu,a}\equiv g^{\nu,\lambda}e^{a}_{\lambda}$ and  $e_{a,\nu}\equiv \delta_{a,b} e^{b}_{\nu}$

For our case we have  a two component the  spin connection $\omega_{x}^{1 2}$ and  $\omega_{y}^{1 2}$ 
\begin{eqnarray}
&&\omega_{x}^{1 2}=\frac{1}{2}e^{\nu,1}(\partial_{x}e^{2}_{\nu}-\partial_{\nu}e^{2}_{x})
-\frac{1}{2}e^{\nu,2}(\partial_{x}e^{1}_{\nu}-\partial_{\nu}e^{1}_{x})
-\frac{1}{2}e^{\rho,a}e^{\sigma,b}(\partial_{\rho}e_{\sigma,c}-\partial_{\sigma}e_{\rho,c})e^{c}_{x};\nonumber\\&&
\omega_{y}^{1 2}=\frac{1}{2}e^{\nu,1}(\partial_{y}e^{2}_{\nu}-\partial_{\nu}e^{2}_{y})
-\frac{1}{2}e^{\nu,2}(\partial_{y}e^{1}_{\nu}-\partial_{\nu}e^{1}_{y})
-\frac{1}{2}e^{\rho,a}e^{\sigma,b}(\partial_{\rho}e_{\sigma,c}-\partial_{\sigma}e_{\rho,c})e^{c}_{y}\nonumber\\&&
\end{eqnarray}
These equations are  further simplified with the help of  equations $(11-13)$  with $e^{1}_{y}=0$ , $e^{1}_{x}=1$ and the Burger    tensor  $\partial_{x}e^{2}_{y}-\partial_{y}e^{2}_{x}=B^{(2)}\delta^{2}(\vec{r})$ . 
\begin{eqnarray}
&&\omega_{x}^{1 2}=\frac{1}{2}g^{\nu,\lambda}e^{1}_{\lambda}(\partial_{x}e^{2}_{\nu}-\partial_{\nu}e^{2}_{x})-\frac{1}{2}g^{\rho,r}e^{1}_{r}g^{\rho,s}e^{2}_{s}[\partial_{\rho}( \delta_{c,b}e^{b}_{\sigma})-\partial_{\sigma}( \delta_{c,d}e^{d}_{\rho})]e^{c}_{x}=\nonumber\\&&
\frac{1}{2}B^{(2)}\delta^{(2)}(\vec{x})[g^{y,x}e^{1}_{x}+g^{y,y}e^{1}_{y}-(g^{x,r}g^{y,s}-
g^{y,r}g^{x,s})(e^{1}_{r}e^{2}_{s}e^{2}_{x}]=\nonumber\\&&
\frac{1}{2}B^{(2)}\delta^{(2)}(\vec{x})[g^{y,x}e^{1}_{x}-(g^{x,x}g^{y,y}-
g^{y,x}g^{x,y})e^{1}_{x}e^{2}_{y}e^{2}_{x}]\approx\nonumber\\&&\frac{1}{2}B^{(2)}\delta^{(2)}(\vec{r})[- \frac{B^{(2)}}{2\pi}\frac{y}{y^2+x^2} -(1-(\frac{B^{(2)}}{2\pi}\frac{y}{y^2+x^2})^2)(\frac{B^{(2)}}{2\pi}\frac{y}{x^2+y^2})(1-\frac{B^{(2)}}{2\pi}\frac{x}{x^2+y^2})]\approx \nonumber\\&&
\frac{1}{2}B^{(2)}\delta^{(2)}(\vec{r})[-\frac{B^{(2)}}{2\pi}\frac{2y-x}{y^2+x^2}]\nonumber\\&&
\end{eqnarray}
and
\begin{eqnarray}
&&\omega_{y}^{1 2}=\frac{1}{2}e^{\nu,1}(\partial_{y}e^{2}_{\nu}-\partial_{\nu}e^{2}_{y})
-\frac{1}{2}e^{\nu,2}(\partial_{y}e^{1}_{\nu}-\partial_{\nu}e^{1}_{y})
-\frac{1}{2}e^{\rho,1}e^{\sigma,2}[\partial_{\rho}(\delta_{c,b}e_{\sigma}^{b})-\partial_{\sigma}(\delta_{c,d}e_{\rho}^{d})]e^{c}_{y}=\nonumber\\&&
\frac{1}{2}g^{\nu,\lambda}e^{1}_{\lambda}[\partial_{y}e^{2}_{\nu}-\partial_{\nu}e^{2}_{y}]
-\frac{1}{2}g^{\nu,r}e^{1}_{r}[\partial_{y}e^{1}_{\nu}-\partial_{\nu}e^{1}_{y}]-\frac{1}{2}g^{\rho,r}e^{1}_{r}g^{\sigma,s}e^{2}_{s}[\partial_{\rho}e^{c}_{\sigma}-\partial_{\sigma}e^{c}_{\rho}]e^{c}_{y}=\nonumber\\&&
-\frac{B^{(2)}}{2}\delta^{(2)}(\vec{r})g^{x,\lambda}e^{1}_{\lambda}-\frac{B^{(2)}}{2}\delta^{(2)}(\vec{r})[g^{x,r}g^{y,s}-g^{y,r}g^{x,s}]e^{1}_{r}e^{2}_{s}e^{2}_{y}\approx -\frac{B^{(2)}}{2}\delta^{(2)}(\vec{r}) \nonumber\\&&
%-\frac{B^{(2)}}{2}\delta^{(2)}(\vec{r})[g^{x,x}+(1-(g^{y,x})^2)\cdot(1-\frac{B^{(2)}}{2\pi}(\frac{x}{x^2+y^2})^2]\approx- B^{(2)}\delta^{(2)}(\vec{r})\nonumber\\&&
\end{eqnarray}
%&&\omega_{y}^{1 2}=\frac{1}{2}(\partial_{x}e^{2}_{y}-\partial_{y}e^{2}_{x})[-g^{1,1}e^{1}_{1}-g^{1,2}e^{1}_{2}]=\frac{1}{2}B^{(2)}\delta^{2}(\vec{r})[-g^{1,1}e^{1}_{1}-g^{1,2}e^{1}_{2}]\nonumber\\&&
%\omega_{x}^{1 2}=\frac{1}{2}(\partial_{x}e^{2}_{y}-\partial_{y}e^{2}_{x})[g^{2,1}+g^{2,2}e^{2}_{2}]=\frac{1}{2}B^{(2)}\delta^{2}(\vec{r})[g^{2,1}+g^{2,2}e^{2}_{2}]\nonumber\\&&
%\end{eqnarray}
To first order first  the Burger vector $B^{(2)}$ the spin connections are given by :
$\omega_{x}^{1 2}=-\omega_{x}^{21}\approx 0$   and  $\omega_{y}^{1 2}=- \omega_{y}^{2 1}\approx -\frac{1}{2}B^{(2)}\delta^2(\vec{r})$.
%\begin{equation}
%\omega_{x}^{1 2}=-\omega_{y}^{1 2}\approx \frac{1}{2}B^{(2)}\delta^2(\vec{r})= \frac{1}{2\pi}B^{(2)}\partial_{z}(\frac{1}{\overline{z}})= \frac{1}{2\pi}B^{(2)}\partial_{\overline{z}}(\frac{1}{z})
%\label{torsion}
%\end{equation}

%is computed from the Christoffel connection $\Gamma^{k}_{i,j} \partial_{k}$  and covariant equation $D_{i}V^{j}(\vec{x})=\partial_{i}V^{j}(\vec{x})+\Gamma^{j}_{i,k}  V^{k}$.
%From  $D_{i}V^{j}(\vec{x})$ and  $D_{i}V^{a}(\vec{x})$ using $V^{a}(\vec{x})=  e^{a}_{i}(\vec{x})V^{i}(\vec{x})$ we determine the spin connection  $\omega_{i,b}^{a}$.

%$D_{i}e^{a}_{j}(\vec{x})=\partial_{i}e^{a}_{j}(\vec{x})-\Gamma^{k}_{i,j} e^{a}_{j}(\vec{x})+ \omega_{a,b}^{i}e^{b}_{j}(\vec{x})=0$

%\vspace{0.1 in}

%This allows to rewrite the Dirac equation  $h=\gamma^{a}(-i\partial_{a}) $ as :

%\begin{equation}
%h^{edge}=e^{\mu}_{a}\cdot\gamma^{a}(-i)(\partial_{\mu}+\frac{1}{8}\omega_{\mu}^{a b}
%[\gamma_{a},\gamma_{b}])
%\label{equation}
%\end{equation}

\vspace{0.2 in}

\textbf{ Appendix-B}

\vspace{0.2 in}

The spinor field operator $\Psi(z,\overline{z};t)$ is decomposed into a sum with different localization contours $n$, $\Psi(z,\overline{z};t)=\sum_{n=0,\pm1,..}\Psi_{n}(z,\overline{z};t)$.The  energy levels for $n\neq0$ are controlled by the  the inverse of the Burger vector .

 We will consider  the  case $n=0$.
For   $\mu=0$,  we have  the expansion in terms of the  eigenspinor  $u_{\epsilon}(s,u)$ for electrons  with $positive $ $chirality$   $v_{\epsilon}(s,u)$  for holes   with $negative$ chirality.

\begin{eqnarray}
&& \int\,ds\int\,du \sqrt{G(s,u)}u^{\dagger}_{\epsilon}(s,u)u_{\epsilon'}(s,u)=\delta(\epsilon,\epsilon')\nonumber\\&&  \int\,ds\int\,du  \sqrt{G(s,u)} v^{\dagger}_{\epsilon}(s,u)v_{\epsilon'}(s,u)=\delta(\epsilon,\epsilon')\nonumber\\&& \sum_{\epsilon >0}u^{\dagger}_{\epsilon}(s,u)u_{\epsilon}(s',u')+ \sum_{\epsilon<0}v^{\dagger}_{\epsilon}(s,u)v_{\epsilon}(s',u')=\frac{\delta^{2}(s-s',u-u')} {\sqrt{G(s,u)}}\nonumber\\&&
\end{eqnarray}

Using the eigenfunctions  $u_{\epsilon}(s,u)$,  $v_{\epsilon}(s,u)$, we  construct the   field operator  $\Psi_{n=0}(s,u;t)$ as a  superposing of particles and holes.   
\begin{eqnarray}
&&\Psi_{n=0}(s,u;t)=\nonumber\\&&\sum_{\epsilon>0} c_{\epsilon}u_{\epsilon}(s,u)e^{-i\frac{\epsilon}{\hbar}t}+\sum_{\epsilon<0}b^{\dagger}_{-\epsilon}v_{\epsilon}(s,u)e^{-i\frac{\epsilon}{\hbar}t}=
\sum_{\epsilon>0}[ c_{\epsilon}u_{\epsilon}(s,u)e^{-i\frac{\epsilon}{\hbar}t}+b^{\dagger}_{\epsilon}v_{-\epsilon}(s,u)e^{i\frac{\epsilon}{\hbar}t}]\nonumber\\&&
\end{eqnarray}
where  $ c_{\epsilon}$ is the annihilation operator for particles with energy $\epsilon$ and $b^{\dagger}_{-\epsilon}$ is the creation operator for a hole with energy $-\epsilon$. $c_{\epsilon}$ and $b_{-\epsilon}$ annihilates the ground state $|0>$ ,$c_{\epsilon}|0>=b_{-\epsilon}|0>=0$. The operators obey anti-commutation relations: $[c_{\epsilon},c^{\dagger}_{\epsilon'}]_{+}=[b_{-\epsilon},b^{\dagger}_{-\epsilon'}]_{+}=\delta(\epsilon,\epsilon')$.

\vspace{0.2 in}

The material properties of the topological insulators are such that the Fermi energy $\mu$ is positive . As a result we  have  electrons and holes  with $positive$ chirality, and  $deep$ holes   with $negative$ spin chiralities.   The energy  is measured with respect  to the chemical potential  $\mu$, $E\equiv\epsilon-\mu$ .As a result   one obtains  a shifted ground state $|\mu>\equiv|\widetilde{0}>$ .

We define new operators using the holes operators  $b_{\epsilon;\pm}$ and electron operator  $c_{\epsilon;+}$:

$c_{E+\mu;+}=\alpha_{E}$ for $E>0$, $c_{E+\mu;+}=\beta^{\dagger}_{-E}$ for $-\mu< E <0$ and $b^{\dagger}_{-(E+\mu);-}=\gamma^{\dagger}_{-E}$ for $-2\mu<E-\mu$.
These operators annihilate the ground state :  $\alpha_{E}|\widetilde{0}>=0$,  $\beta_{-E}|\widetilde{0}>=0$ and $\gamma_{-E}|\widetilde{0}>=0$.

\begin{eqnarray} 
&&\Psi_{n=0}(s,u;t)=
\nonumber\\&&
\sum_{E=0}^{E= E_{\Lambda}} c_{E+\mu;+}u_{E+\mu}(s,u)e^{-i\frac{E}{\hbar}t}+ \sum_{E=-\mu}^{E=0}c_{(E+\mu);+}u_{E+\mu}(s,u)e^{-i\frac{E}{\hbar}t}+ \sum_{-\mu-E_{\Lambda}}^{-\mu}b^{\dagger}_{-(E+\mu);-} v_{E+\mu}(s,u)e^{-i\frac{E}{\hbar}t}\nonumber\\&&=
\sum_{E=0}^{E=E_{\Lambda}}\alpha_{E}u_{E+\mu}(s,u)e^{-i\frac{E}{\hbar}t}+ \sum_{E=-\mu}^{E=0} \beta^{\dagger}_{-E}u_{E+\mu}(s,u)e^{-i\frac{E}{\hbar}t}+\sum_{E=-\mu-E_{\Lambda}}^{E=-\mu} \gamma^{\dagger}_{-E} v_{E+\mu}(s,u)e^{-i\frac{E}{\hbar}t}\nonumber\\&&= \sum_{E=0}^{E=E_{\Lambda}}\alpha_{E}u_{E+\mu}(s,u)e^{-i\frac{E}{\hbar}t}+ \sum_{E=0}^{E=\mu}\beta^{\dagger}_{E}u_{-E+\mu}(s,u)e^{i\frac{E}{\hbar}t}+\sum_{E=\mu+E_{\Lambda}}^{E=\mu}\gamma^{\dagger}_{E}v_{-E+\mu}(s,u)e^{i\frac{E}{\hbar}t}\nonumber\\&&
\end{eqnarray}
where  $E_{\Lambda}$ is  energy the  below the Dirac point .
As a result, the  edge Hamiltonian is given by:
$H^{edge}\approx[\sum_{E=0}^{E=E_{\Lambda}} \alpha^{\dagger}_{E}\alpha_{E} E+ \sum_{E=0}^{E=\mu}\beta^{\dagger}_{E}\beta_{E} E+ \sum_{E=\mu+E_{\Lambda}}^{E=\mu}\gamma^{\dagger}_{E} \gamma_{E} E]$

For  most of the cases, the   chemical  potential  is large, and we  can approximate  the spinor operator  by a sum of states for particles and holes,   ignoring the deep hole band:
 \begin{equation}
\Psi_{n=0}(s,u;t)\approx\sum_{E>0} [\alpha_{E}u_{E+\mu}(s,u)e^{-i\frac{E}{\hbar}t}+\beta^{\dagger}_{E}u_{-E+\mu}(s,u)e^{i\frac{E}{\hbar}t}]
\label{short}
\end{equation}
 
\pagebreak

\vspace{0.2 in}

\textbf{ Appendix-C}

\vspace{0.2 in}

The wave functions are given by:   
\begin{eqnarray}
&&U_{\epsilon(l,Q_{-})\uparrow}(\theta,u)= G^{\frac{-1}{4}}(\theta,u)\cdot[U^{(even,k)}_{\epsilon(l,Q_{-})\uparrow}(\theta,u)+
U^{(odd,k)}_{\epsilon(l,Q_{-})\uparrow}(\theta,u)];\nonumber\\&&
U^{(even,k)}_{\epsilon(l,Q_{-})\uparrow}(\theta,u)=
2i e^{\frac{-i}{2} \kappa(l,Q_{-})}[e^{\zeta(\theta,u)\sin[2\theta]}e^{-i\zeta(\theta,u)(\eta(u)-\cos[2\theta])}e^{il(2\theta+\frac{3\pi}{2})}\sin[Q_{-}u+\frac{1}{2}\kappa(l,Q_{-})]\nonumber\\&&
+(-1)^l e^{-\zeta(\theta,u)\sin[2\theta]}e^{-i\zeta(\theta,u)(-\eta(u)+\cos[2\theta])}e^{-il(2\theta+\frac{3\pi}{2})}\sin[Q_{-}u-\frac{1}{2}\kappa(l,Q_{-})]];\nonumber\\&&
U^{(odd,k)}_{\epsilon(l,Q_{-})\uparrow}(\theta,u)=
2 e^{\frac{-i}{2} \kappa(l,Q_{-})}[e^{\zeta(\theta,u)\sin[2\theta]}e^{-i\zeta(\theta,u)(\eta(u)-\cos[2\theta])}e^{il(2\theta+\frac{3\pi}{2})}\cos[Q_{-}u+\frac{1}{2}\kappa(l,Q_{-})]\nonumber\\&&
+(-1)^l e^{-\zeta(\theta,u)\sin[2\theta]}e^{-i\zeta(\theta,u)(-\eta(u)+\cos[2\theta])}e^{-il(2\theta+\frac{3\pi}{2})}\cos[Q_{-}u-\frac{1}{2}\kappa(l,Q_{-})]];\nonumber\\&&
U_{\epsilon(l,Q_{-})\downarrow}(\theta,u)=G^{\frac{-1}{4}}(\theta,u)\cdot[U^{(even,k)}_{\epsilon(l,Q_{-})\downarrow}(\theta,u)+
U^{(odd,k)}_{\epsilon(l,Q_{-})\uparrow}(\theta,u)];\nonumber\\&&
U^{(even,k)})_{\epsilon(l,Q_{-})\downarrow}(\theta,u)=
2i e^{\frac{-i}{2} \kappa(l,Q_{-})}[e^{\zeta(\theta,u)\sin[2\theta]}e^{i\zeta(\theta,u)(\eta(u)-\cos[2\theta])}e^{i((l+1)(2\theta+\frac{3\pi}{2}))}\sin[Q_{-}u-\frac{1}{2}\kappa(l,Q_{-})]\nonumber\\&&
-(-1)^l e^{-\zeta(\theta,u)\sin[2\theta]}e^{i\zeta(\theta,u)(-\eta(u)+\cos[2\theta])}e^{-i((l+1)(2\theta+\frac{3\pi}{2}))}\sin[Q_{-}u+\frac{1}{2}\kappa(l,Q_{-})]];\nonumber\\&&
U^{(odd,k)}_{\epsilon(l,Q_{-})\downarrow}(\theta,u)=2e^{\frac{-i}{2} \kappa(l,Q_{-})}[e^{\zeta(\theta,u)\sin[2\theta]}e^{i\zeta(\theta,u)(\eta(u)-\cos[2\theta])}e^{i((l+1)(2\theta+\frac{3\pi}{2}))}\cos[Q_{-}u-\frac{1}{2}\kappa(l,Q_{-})]\nonumber\\&&
-(-1)^l e^{-\zeta(\theta,u)\sin[2\theta]}e^{i\zeta(\theta,u)(-\eta(u)+\cos[2\theta])}e^{-i((l+1)(2\theta+\frac{3\pi}{2}))}\cos[Q_{-}u+\frac{1}{2}\kappa(l,Q_{-})]];\nonumber\\&&
\end{eqnarray} 
Similarly for the second pair we obtain the wave function:
\begin{eqnarray}
&&U_{\epsilon(l,Q_{+})\uparrow}(\theta,u)= G^{\frac{-1}{4}}(\theta,u)\cdot[U^{(even,k)}_{\epsilon(l,Q_{+})\uparrow}(\theta,u)+
U^{(odd,k)}_{\epsilon(l,Q_{+})\uparrow}(\theta,u)];\nonumber\\&&
U^{(even,k)}_{\epsilon(l,Q_{+})\uparrow}(\theta,u)=
2i e^{\frac{-i}{2} \kappa(l,Q_{+})}[(-1)^{l}e^{-\zeta(\theta,u)\sin[2\theta]}e^{-i\zeta(\theta,u)(-\eta(u)+\cos[2\theta])}e^{il(2\theta+\frac{3\pi}{2})}\sin[Q_{+}u+\frac{1}{2}\kappa(l,Q_{+})]\nonumber\\&&
+ e^{\zeta(\theta,u)\sin[2\theta]}e^{-i\zeta(\theta,u)(\eta(u)-\cos[2\theta])}e^{-il(2\theta+\frac{3\pi}{2})}\sin[Q_{+}u-\frac{1}{2}\kappa(l,Q_{+})]];\nonumber\\&&
U^{(odd,k)}_{\epsilon(l,Q_{+})\uparrow}(\theta,u)=
2  e^{\frac{-i}{2} \kappa(l,Q_{+})}[(-1)^{l}e^{-\zeta(\theta,u)\sin[2\theta]}e^{-i\zeta(\theta,u)(-\eta(u)+\cos[2\theta])}e^{il(2\theta+\frac{3\pi}{2})}\cos[Q_{+}u+\frac{1}{2}\kappa(l,Q_{+})]\nonumber\\&&
+ e^{\zeta(\theta,u)\sin[2\theta]}e^{-i\zeta(\theta,u)(\eta(u)-\cos[2\theta])}e^{-il(2\theta+\frac{3\pi}{2})}\cos[Q_{+}u-\frac{1}{2}\kappa(l,Q_{+})]]; \nonumber\\&&    
U_{\epsilon(l,Q_{+})\downarrow}(\theta,u)= G^{\frac{-1}{4}}(\theta,u)\cdot[U^{(even,k)}_{\epsilon(l,Q_{+})\downarrow}(\theta,u)+
U^{(odd,k)}_{\epsilon(l,Q_{+})\downarrow}(\theta,u)];\nonumber\\&&
U^{(even,k)}_{\epsilon(l,Q_{+})\downarrow}(\theta,u)=
2i e^{\frac{-i}{2} \kappa(l,Q_{+})}[-(-1)^{l}e^{-\zeta(\theta,u)\sin[2\theta]}e^{i\zeta(\theta,u)(-\eta(u)+\cos[2\theta])}e^{i(l+1)(2\theta+\frac{3\pi}{2})}\sin[Q_{+}u-\frac{1}{2}\kappa(l,Q_{+})]\nonumber\\&&
+ e^{\zeta(\theta,u)\sin[2\theta]}e^{i\zeta(\theta,u)(\eta(u)-\cos[2\theta])}e^{-i(l+1)(2\theta+\frac{3\pi}{2})}\sin[Q_{+}u+\frac{1}{2}\kappa(l,Q_{+})]];\nonumber\\&&
U^{(odd,k)}_{\epsilon(l,Q_{+})\downarrow}(\theta,u)=
2 e^{\frac{-i}{2} \kappa(l,Q_{+})}[-(-1)^{l}e^{-\zeta(\theta,u)\sin[2\theta]}e^{i\zeta(\theta,u)(-\eta(u)+\cos[2\theta])}e^{i(l+1)(2\theta+\frac{3\pi}{2})}\cos[Q_{+}u-\frac{1}{2}\kappa(l,Q_{+})]\nonumber\\&&
+ e^{\zeta(\theta,u)\sin[2\theta]}e^{i\zeta(\theta,u)(\eta(u)-\cos[2\theta])}e^{-i(l+1)(2\theta+\frac{3\pi}{2})}\cos[Q_{+}u+\frac{1}{2}\kappa(l,Q_{+})]];\nonumber\\&&    
\end{eqnarray}
where $G^{\frac{-1}{4}}(\theta,u)$  is  the Jacobian transformation induced by the metric tensor.

%We replace the basis  $u^{(n;k,l)}(s,u)$ by  the orthogonal basis $\alpha^{(n);k,l)}(s,u)$ defined in the   following way:
%$ \alpha^{(n;k=1,l=1)}(s,u)=u^{(n;k=1,l=1)}(s,u)$  and  $ \alpha^{(n;k=m,l=m')}(s,u)$ is given by:
%\begin{eqnarray}
%&&\alpha^{(n;k=m,l=m')}(s,u)=\frac{1}{\sqrt{N_{m,m'}}}[u^{(n;m,m')}(s,u)-\sum_{j=1}^{m-1} \sum_{j'=1}^{m'-1}(\alpha^{(n;j,j')}|u^{(n;m,m')})\alpha^{(n;j,j')}(s,u)]\nonumber\\&&
%N_{m,m'}=(u^{(n;m,m')}|u^{(n);m,m')})-\sum_{j=1}^{m-1} \sum_{j'=1}^{m'-1}(\alpha^{(n;j,j')}|u^{(n;m,m')}|^{2} u^{(n;m,m')}(s,u)\nonumber\\&&
%\end{eqnarray}
%where $(\alpha^{(n;j,j')}|u^{(n;m,m')})\equiv\int\,ds \int\,du \sqrt{G}(\alpha^{(n;j,j')}(s,u))^{*}u^{(n;m,m')}(s,u)$  represent the scalar product . 

%\vspace{0.2 in}

\pagebreak

\clearpage
\begin{figure}
\begin{center}
\includegraphics[width=6.5 in ]{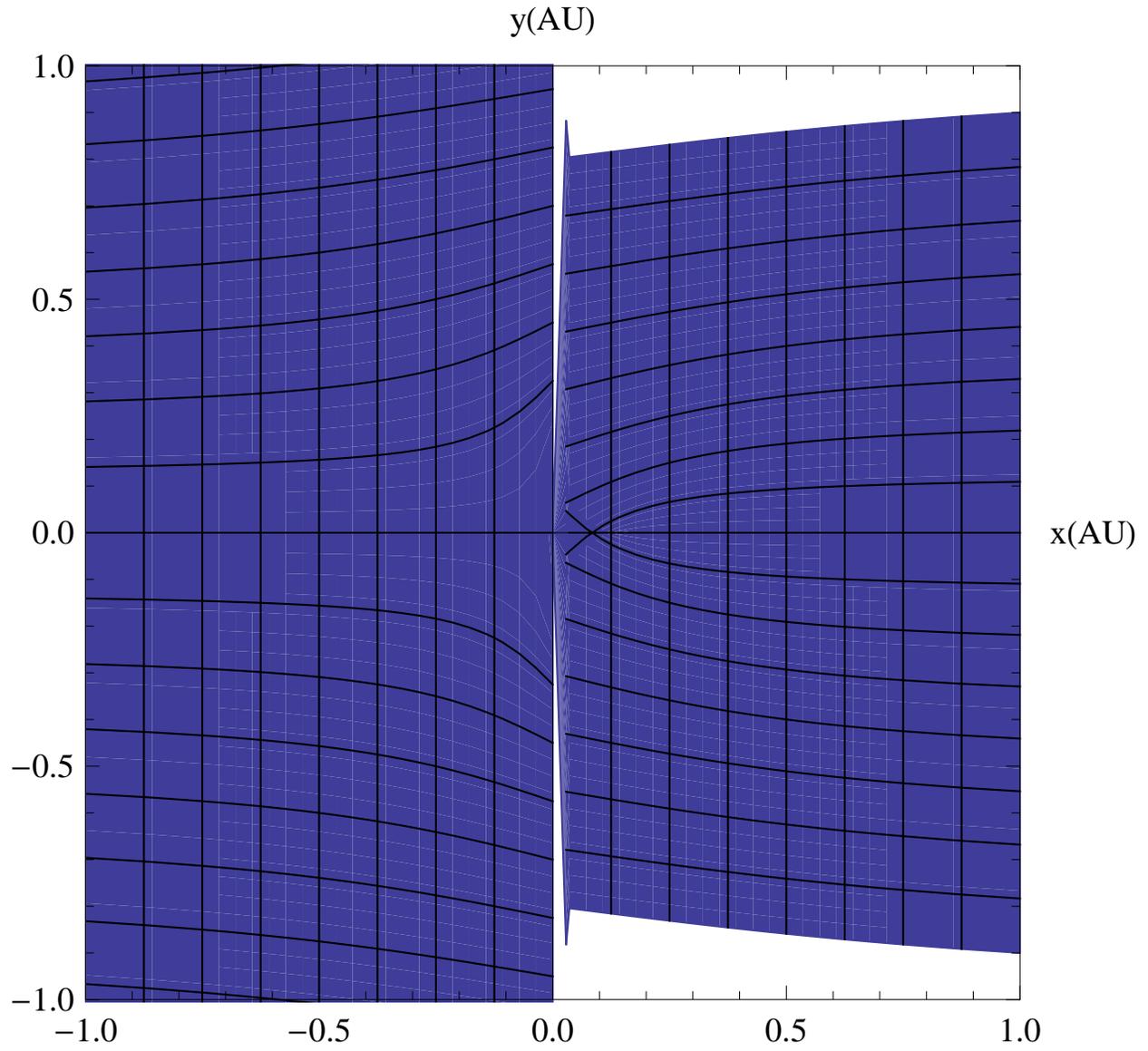}
\end{center}
\caption{The edge dislocation with the core at $x=y=0$,  modifies the coordinate $x$,$y$ to   $X$, $Y$  in the presence of the edge dislocation   with the  Burger vector $ B^{(2)}$}  
\end{figure}

\clearpage
\begin{figure}
\begin{center}
\includegraphics[width=7.0 in ]{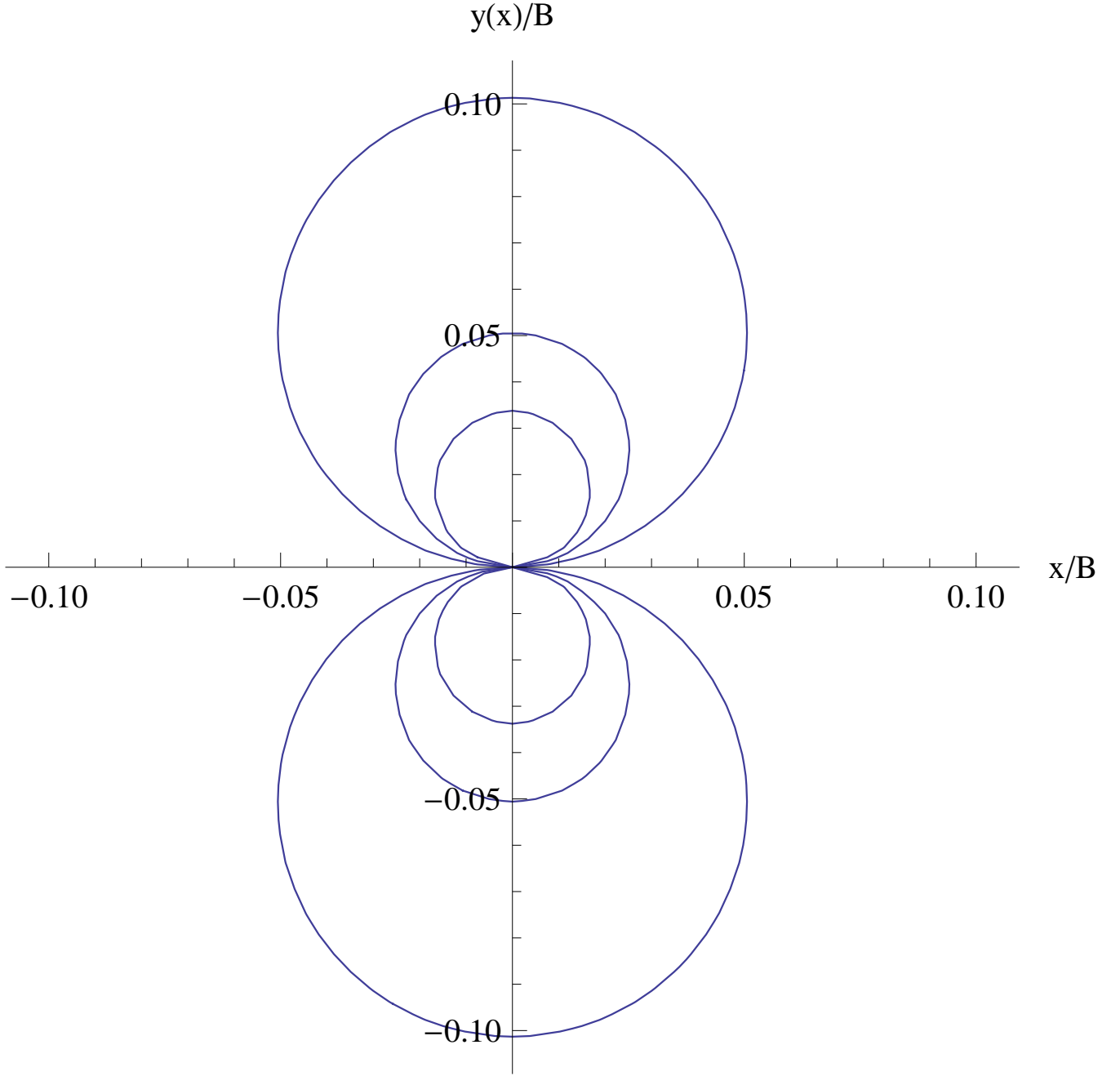}
\end{center}
\caption{The contours  $(x(s))^2+ (y(s)-\frac{R_{g}}{ n})^2 =(\frac{R_{g}}{n})^2$ for $n=\pm1,\pm2,\pm3$(in  decreasing  size ),$R_{g}(n)=\frac{R_{g}}{n}$. $n=0$ corresponds to the equation $y(s)=0$ and $|y|>d $ (see the text).  The  the distance is measured in  units  of the Burger vector $B^{(2)}$.}   
\end{figure}

%\clearpage

%\begin{figure}
%\begin{center}
%\includegraphics[width=7.4 in ]{Nabaro.eps}
%\end{center}
%\caption{The}
%\end{figure}

\clearpage

\begin{figure}
\begin{center}
\includegraphics[width=7.4 in ]{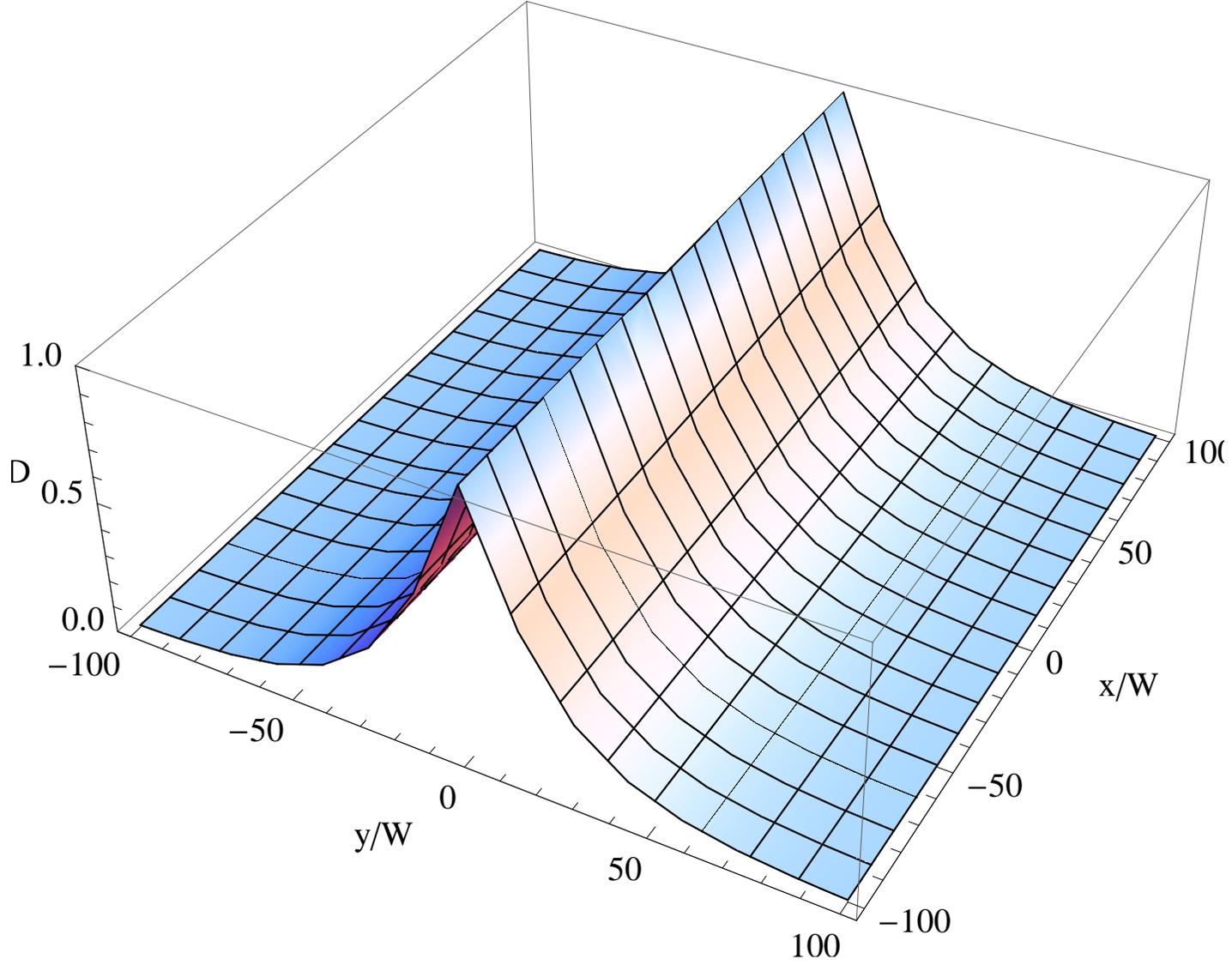}
\end{center}
\caption{The  tunenling  density of states for $\hat{D}^{Peierls}(\frac{x}{\widetilde{W}},\frac{y}{\widetilde{W}})$ for the Peierls dislocation confined to the line $|x|< \widetilde{W}$ for the STM  voltage $V$, $\mu+eV<0 $}   
\end{figure}

%\clearpage
%\begin{figure}
%\begin{center}
%\includegraphics[width=7.0 in ]{ndensity_10.eps}
%\end{center}
%\caption{The  tunneling density   of states  for  $n=0$  ,  $\frac{dI}{dV}\propto D^{(n=0)}(\mu=120mV,V) $}
%\end{figure}

\clearpage
\begin{figure}
\begin{center}
\includegraphics[width=7.0 in ]{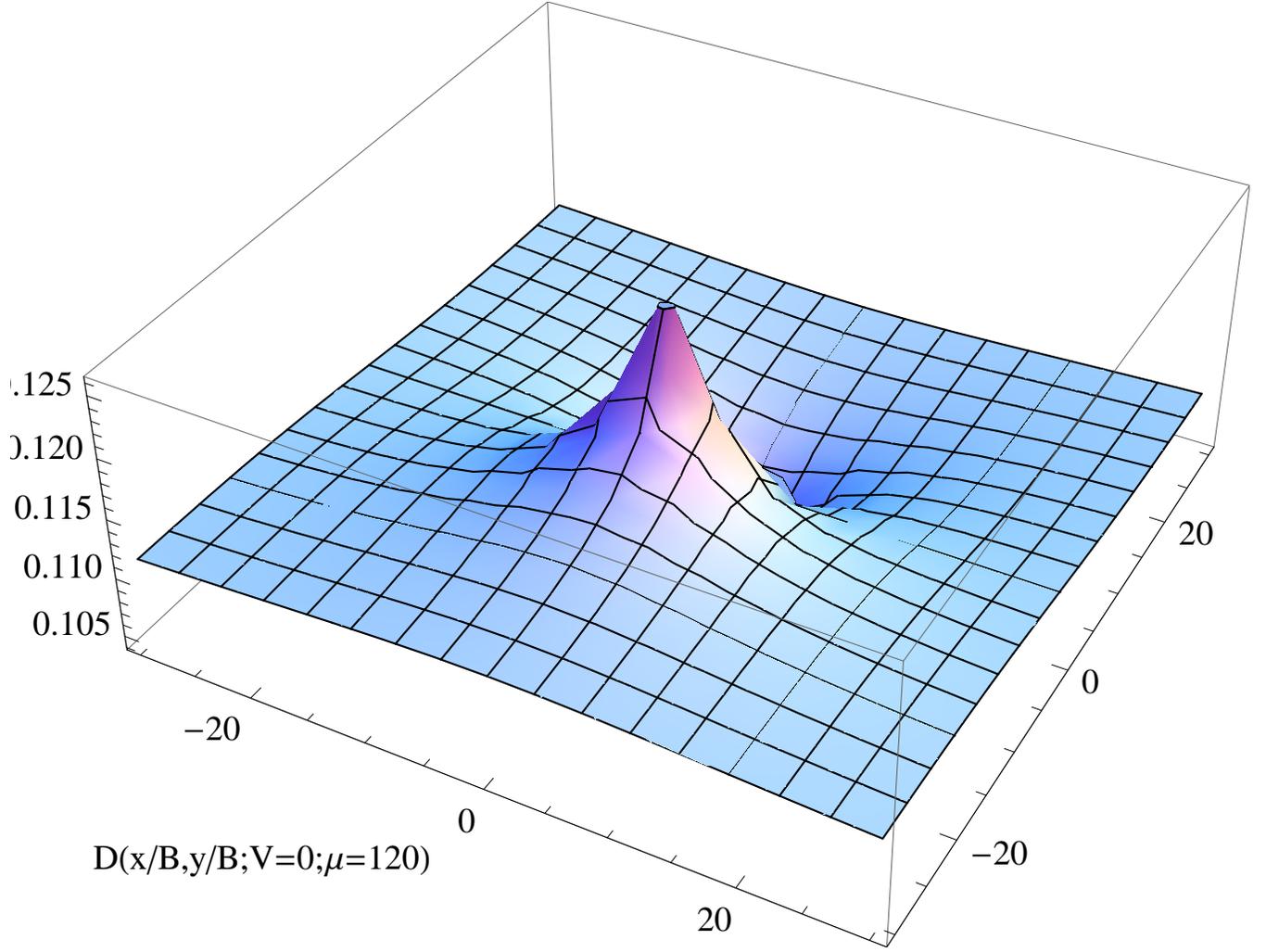}
\end{center}
\caption{The  tunneling density   of states  for  $n=0$  ,  $\frac{dI}{dV}\propto D^{(n=0)}(\frac{x}{B^{(2)}},\frac{y}{B^{(2)}};\mu=120mV) $.  The right corner  represents the intersection of the $x$ coordinate which runs from $30$ (right corner)  to $-30$  and the $y$ coordinate  which runs from $-30$ (right corner)  to $30$ in units of the Burger vector.}
\end{figure}

\clearpage

\begin{figure}
\begin{center}
\includegraphics[width=7.0 in ]{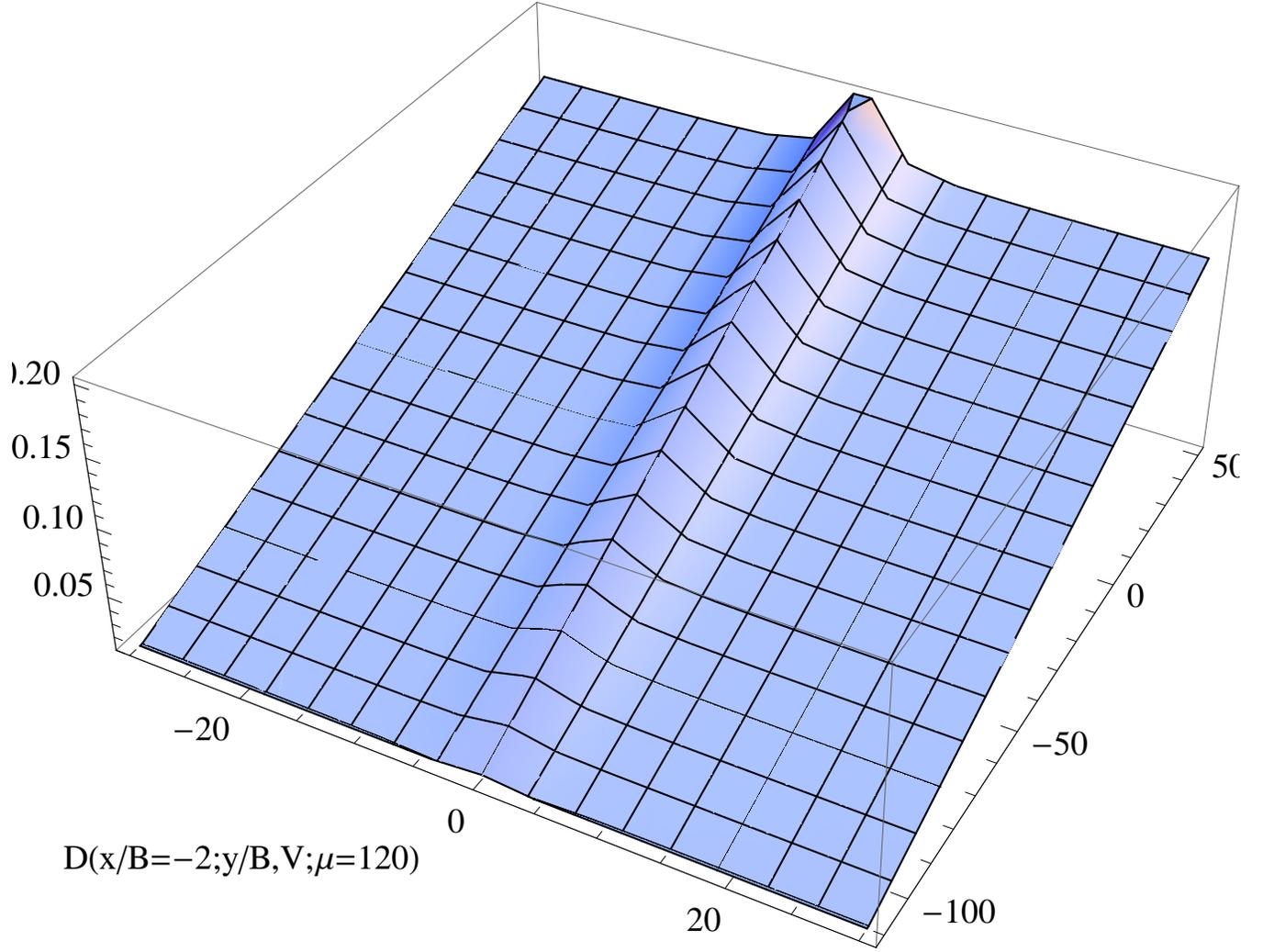}
\end{center}
\caption{The  tunneling density   of states  for  $n=0$  as a function  of $y$ and $V$ $\frac{dI}{dV}\propto D^{(n=0)}(\frac{x}{B^{(2)}}=-2,\frac{y}{B^{(2)}};\mu=120mV) $. The voltage range is $-120 \leq V  \leq  50$ and the $y$ coordinate is in the range   $-30 \leq \frac{y}{B^{(2)}}  \leq  30$.}
\end{figure}

\clearpage
\clearpage
\begin{figure}
\begin{center}
\includegraphics[width=7.0 in ]{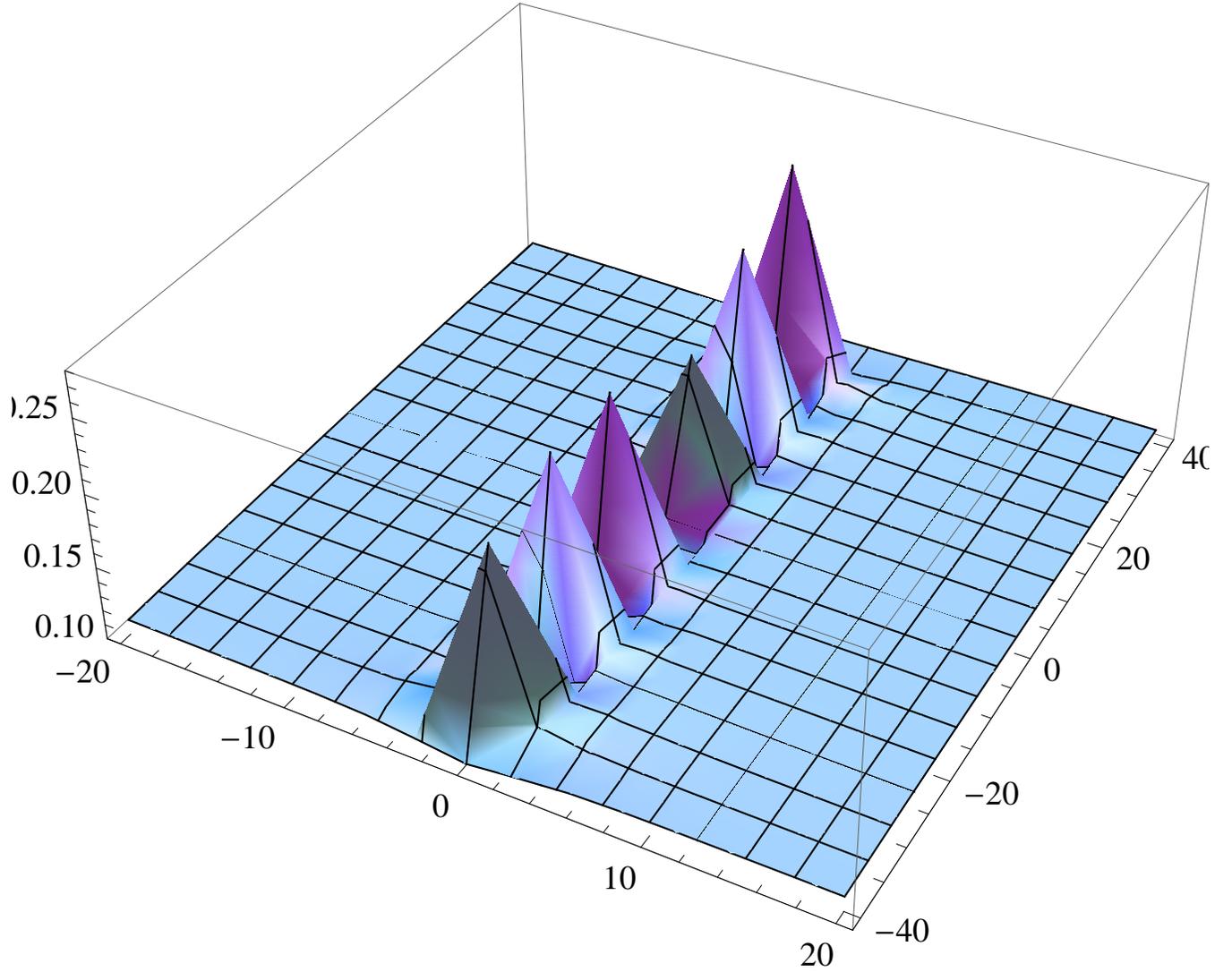}
\end{center}
\caption{Many Dislocations - with the core of the dislocations at   $[x_{w},y=0]$ , $w=1,2...2M$;   The maximum of the tunneling density of states  is confined along  $y=0$. The coordinates of the tunneling density of states are restricted to :  $-40 \leq \frac{x}{B^{(2)}} \leq  40$ and  $-20 \leq \frac{y}{B^{(2)}} \leq  20$. }  
\end{figure}
\clearpage

\clearpage
\begin{figure}
\begin{center}
\includegraphics[width=7.0 in ]{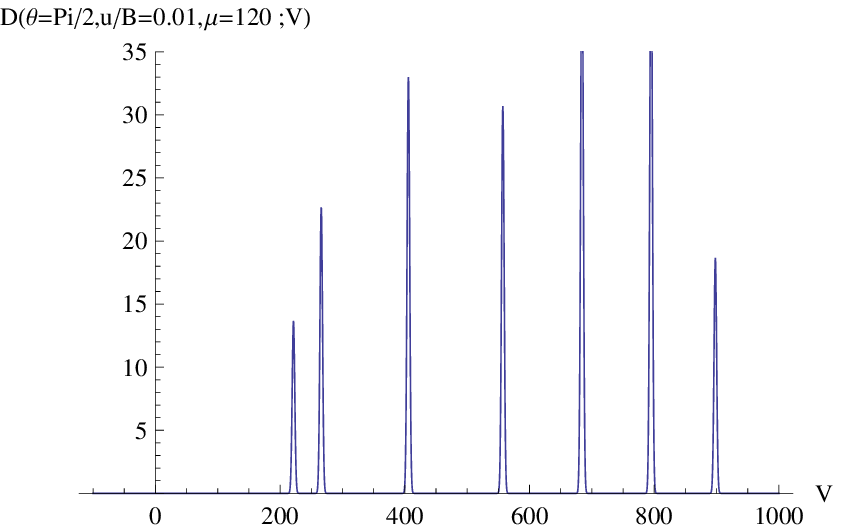}
\end{center}
\caption{The  discrete tunneling density  of states  for  $n=1$, as a function of the voltage $V$ $D^{(n=1)}(V;\theta =\frac{\pi}{2},\frac{u}{B^{(2)}},\mu=120mV)$}
\end{figure}
\clearpage

\begin{figure}
\begin{center}
\includegraphics[width=7.0 in ]{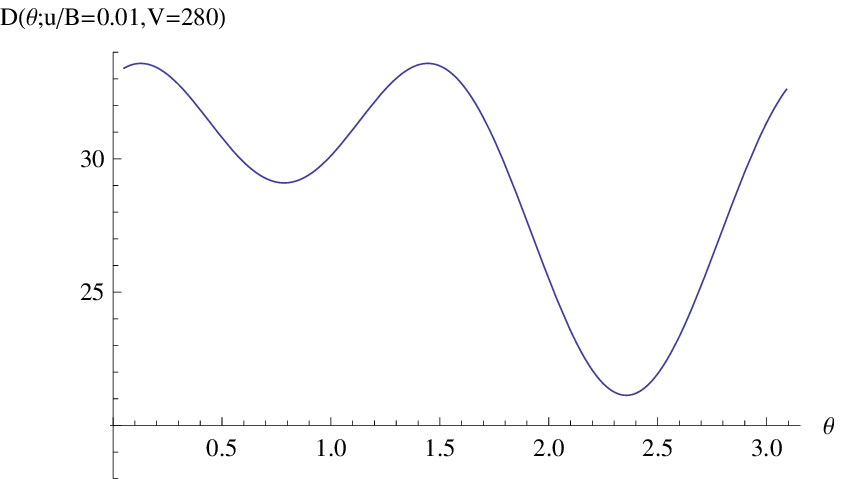}
\end{center}
\caption{The tunneling density of states as a function of $\theta$  $D^{(n=1)}(\theta; \frac{u}{B^{(2)}}=0.01, V=280 mV,\mu=120mV)$}  
\end{figure}

\clearpage

\clearpage
\begin{figure}
\begin{center}
\includegraphics[width=7.0 in ]{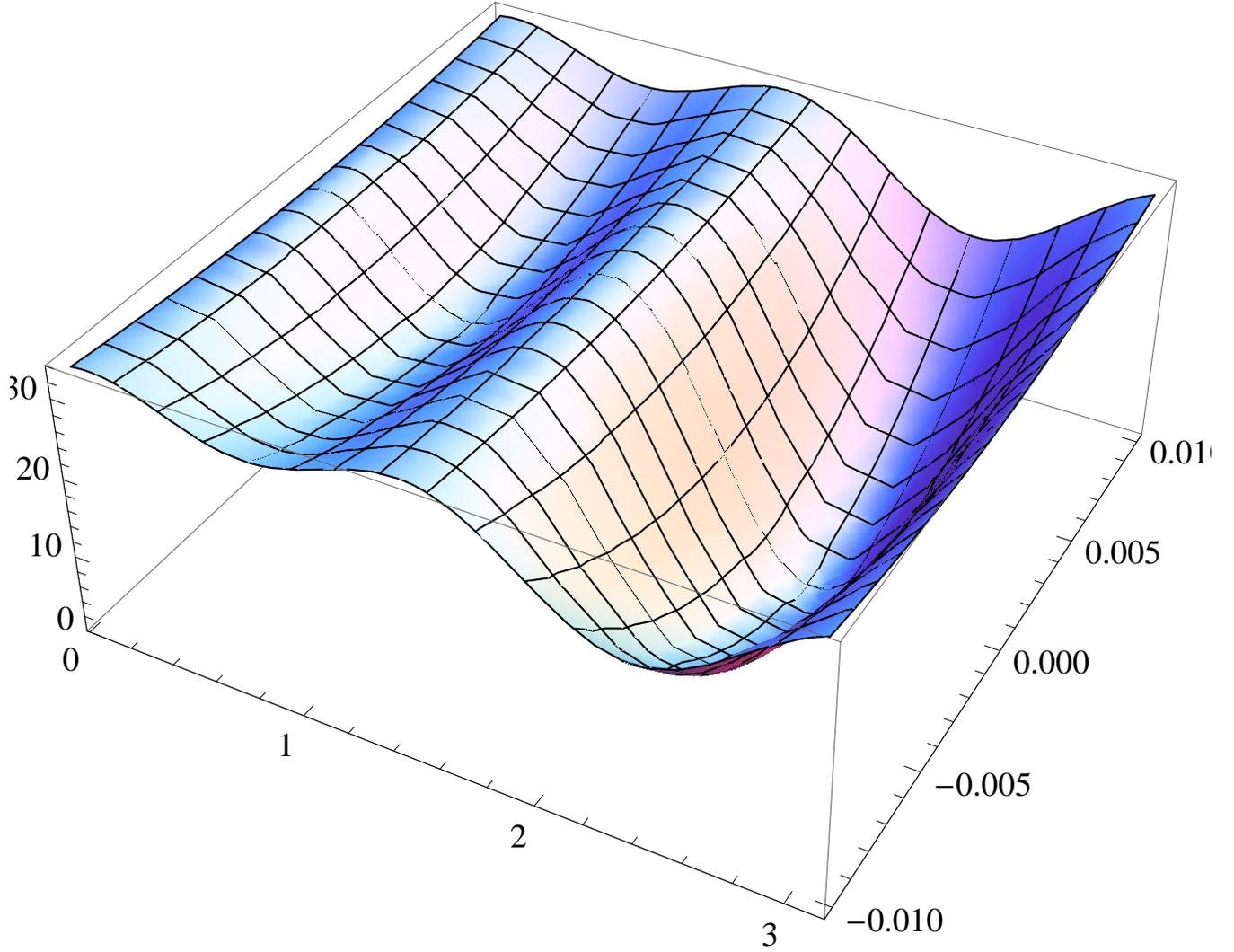}
\end{center}
\caption{The tunneling  density of states as a function of  $\theta$ and $u$ at a fixed voltage $V=280mV$   $D^{(n=1)}(\theta, \frac{u}{B^{(2)}}; V=280 mV,\mu=120mV)$}  
\end{figure}
\clearpage

%\vspace{0.1 in}

\pagebreak

%\textbf{Plan of operation}

\end{document}